\documentclass[widetext,showpacs,preprintnumbers,superscriptaddress,nofootinbib]{revtex4}

\usepackage[dvipdfm]{graphicx}
\usepackage{amssymb}
\usepackage{amsmath}
\usepackage{dcolumn}
\usepackage{bm}
\usepackage{natbib}
\usepackage{multirow}
\usepackage{subfigure}

\begin{document}

\preprint{RIKEN-MP-76,\ KEK-TH-1663}

\title{Effective gravitational interactions of dark matter axions}

\author{Toshifumi Noumi}
\email{toshifumi.noumi@riken.jp}
\affiliation{Mathematical Physics Laboratory, RIKEN Nishina Center,
Saitama 351-0198, Japan}

\author{Ken'ichi Saikawa}
\email{saikawa@th.phys.titech.ac.jp}
\affiliation{Department of Physics, Tokyo Institute of Technology,
2-12-1 Ookayama, Meguro-ku, Tokyo 152-8551, Japan}

\author{Ryosuke Sato}
\email{rsato@post.kek.jp}
\affiliation{Theory Center, High Energy Accelerator Research Organization (KEK),
1-1 Oho, Tsukuba, Ibaraki 305-0801, Japan}

\author{Masahide Yamaguchi}
\email{gucci@phys.titech.ac.jp}
\affiliation{Department of Physics, Tokyo Institute of Technology,
2-12-1 Ookayama, Meguro-ku, Tokyo 152-8551, Japan}
 
\date{\today}

\begin{abstract}
We investigate the structure of gravitational self-interactions of coherently oscillating axions in the general relativistic framework.
A generic action
for a massive scalar field in the Friedmann-Robertson-Walker background
is first introduced
based on the effective field theory
approach to cosmological perturbations.
Using the obtained setup,
we evaluate the effective
gravitational interaction
of the massive scalar field,
i.e. scalar quartic interactions
mediated by metric perturbations.
Applying the results to the system of dark matter axions, we estimate their self-interaction rate and discuss its implications for
the axion Bose-Einstein condensate dark matter scenario.
Leading contributions for the gravitational interactions of axions are given by the process mediated by the dynamical graviton field,
which is essentially the Newtonian potential induced by
fluctuations of the background fluids.
We find that it leads to the same order of magnitude for the interaction rate of dark matter axions in the condensed regime,
compared with the results of previous studies using the Newtonian approximation.
\end{abstract}

\pacs{14.80.Va,\ 95.35.+d,\ 98.80.Cq}

\maketitle

\begin{widetext}
\tableofcontents\vspace{5mm}
\end{widetext}

\section{\label{sec1}Introduction}
Recent developments of observational studies have constrained the properties of dark matter significantly, yet its origin is unknown.
Since the standard model of particle physics cannot explain the origin of dark matter,
it is expected that physics beyond the standard model should give rise to some explanation.
So far, the weakly interacting massive particle (WIMP) has received a lot of attention as a candidate of particle dark matter,
since it naturally explains the observed abundance of dark matter if it has a weak scale mass~\cite{Jungman:1995df}.
The existence of such a particle is suggested by some new physics solving the hierarchy problem of the standard model,
because they predict new particles with weak scale masses. The most representative example is supersymmetry \cite{Martin:1997ns}.
However, the null observation of the new physics at the LHC experiment tells us that using new physics as a solution of the hierarchy problem comes up against difficulties.
Accordingly, the search for alternatives will become more important. 

The axion~\cite{Weinberg:1977ma} is another leading candidate of dark matter, which emerges out of the solution to
the strong CP problem of quantum chromodynamics (QCD)~\cite{Peccei:1977hh}.
In some respects the properties of dark matter axions are different from those of WIMPs. They are produced nonthermally in the early universe and
described as a coherently oscillating scalar field~\cite{Preskill:1982cy}.
Since this coherent oscillation is interpreted as the highly condensed Bose gas of axions, there is some possibility for dark matter axions
of forming Bose-Einstein condensate (BEC) in the universe.
This possibility has received considerable interest recently, due to the suggestion by Sikivie and Yang~\cite{Sikivie:2009qn}
that gravitational interactions can thermalize the system.

The formation of axion BEC dark matter, if it occurred, leads to some interesting phenomenological implications for astrophysics and cosmology.
It was argued that the angular momentum distribution of infalling dark matter particles affects the structure of inner caustics,
which is the overdense region produced by the fall of dark matter surrounding the galaxy~\cite{Sikivie:1999jv}.
If the particles have a net overall rotation, which is predicted by axion BEC dark matter~\cite{Sikivie:2010bq}, the inner caustics have a ringlike structure.
Such a structure is not predicted
for WIMPs, where infalling particles are irrotational,
and hence there is a possibility of distinguishing dark matter candidates on observational grounds.
Furthermore, there is another motivation to investigate the thermalization process of dark matter axions, after the suggestion in Ref.~\cite{Erken:2011vv} that
some cosmological parameters such as the effective number $N_{\rm eff}$ of neutrinos and the baryon-to-photon ratio are
modified if axions have thermal contact with other species by means of gravitational interactions.
The predicted value for $N_{\rm eff}$ significantly conflicts with the observed value $N_{\rm eff}\simeq 3\mathchar`-4$~\cite{Ade:2013zuv}.

The crucial point for the scenarios discussed above is that the thermalization of the system occurs due to gravitational interactions.
Gravitational thermalization of dark matter axions was first discussed in detail by the authors of Ref.~\cite{Erken:2011dz},
and they claimed
that the formation of axion BEC occurs in the condensed regime, where the interaction rate is large
compared to the typical energy exchanged in the interaction.
In that regime, the interaction rate is given by first-order terms in the coupling constant, which is greater than
the usual kinetic estimation given by second-order terms.

The thermalization process in the condensed regime was further studied by two of the present authors in Ref.~\cite{Saikawa:2012uk}
by developing the formalism to compute the expectation value of the occupation number of axions.
In Ref.~\cite{Saikawa:2012uk}, it was shown that if coherently oscillating axions are represented as coherent states,
they do not have thermal contact with other particle species represented as number states.
This result implies that the concerns about the effects on cosmological parameters such as $N_{\rm eff}$ are fictitious.
On the other hand, the analysis in Ref.~\cite{Saikawa:2012uk} led to the same result as Ref.~\cite{Erken:2011dz} for the gravitational self-interaction rate of axions:
\begin{equation}
\Gamma \simeq \frac{4\pi Gm^2n}{(\delta p)^2}, \label{eq1-1}
\end{equation}
where $G$ is Newton's constant, $m$ is the mass of the axion, $n$ is its number density, and $\delta p$ is its momentum dispersion.
Since $n\propto a^{-3}$ and $\delta p\propto a^{-1}$, this rate scales as $\Gamma \propto a^{-1}$, where $a$ is the scale factor of the universe.
Then, it exceeds the expansion rate $H\sim 1/t$ when the temperature of the universe becomes $T\simeq\mathcal{O}(10^2\mathchar`-10^3)$eV,
which might imply that
the gravitational self-interactions of axions affect the evolution of dark matter axions
and the occupation number of axions changes rapidly at that time.
However, there are several issues that need further investigation.
First, the interaction rate in Eq.~\eqref{eq1-1} seems not to be applicable to the modes outside the horizon ($\delta p \lesssim H$),
since the expression~\eqref{eq1-1} was derived based on the Newtonian approximation for the interaction Hamiltonian of the gravitational coupling in the flat Minkowski background.
It is unclear whether the previous result [Eq.~\eqref{eq1-1}] is modified if we include the correction coming from general relativity.
Second, even if the estimation for the interaction rate in Eq.~\eqref{eq1-1} turns out to be correct in the expanding universe,
it is not fully understood how the system evolves after the interaction rate exceeds the expansion rate.

It should be emphasized that consideration on the second question described above is more subtle than the first one. In the previous studies~\cite{Erken:2011dz,Saikawa:2012uk},
the rate $\Gamma$ shown in Eq.~\eqref{eq1-1} was simply called the ``thermalization rate"
with the assumption that the system develops toward thermal equilibrium after the gravitational self-interactions become non-negligible.
However, the completion of the gravitational thermalization and the formation of axion BEC were not explicitly shown in Refs.~\cite{Erken:2011dz,Saikawa:2012uk}.
To show the completion of the thermalization, it is necessary to confirm that almost all axions transit into the lowest energy state,
which might imply that the axion field becomes homogenized in the real space.
Recently, the occurrence of gravitational thermalization was doubted by the authors of Ref.~\cite{Davidson:2013aba},
in which it is claimed that the homogenization of the classical axion field seems to conflict with
the naive argument based on the linearized classical Einstein gravity, 
where the inhomogeneous density fluctuations in the axion field grow. 
Regarding these nontrivial issues, in this paper we just call $\Gamma$ the interaction rate
and concentrate on the first question, the general relativistic correction to the estimation of $\Gamma$.

The purpose of this paper is to investigate all possible elementary processes contributing to the gravitational self-interactions of axions
in the expanding universe, and to reanalyze the interaction rate $\Gamma$.
Here we use the formalism based on the quantum field theory developed in Ref.~\cite{Saikawa:2012uk}, since
in the quantum analysis the structure of interactions involving the creation and annihilation of particles can be investigated straightforwardly.
Indeed, it is possible to write down the schematics of various elementary processes in terms of the diagrams accompanied by axions and gravitons.
Then, we aim to discuss whether the estimation of Eq.~\eqref{eq1-1} remains correct when we include the general relativistic effects, and if so,
what kinds of interactions contribute to the ``fast" process whose rate is estimated by Eq.~\eqref{eq1-1}.
It is expected that there exists a distinction between modes inside and outside the horizon, since the gravitational thermalization should proceed
by means of causal interactions occurring inside the horizon.

\begin{figure}[htbp]
\begin{center}
\includegraphics[scale=0.75]{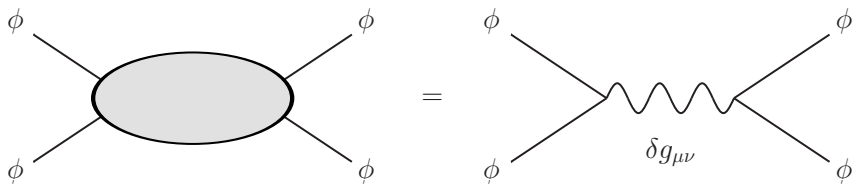}
\end{center}
\caption{Effective quartic interaction mediated by the metric perturbations.}
\label{fig1}
\end{figure}

In the general relativistic framework,
the gravitational interaction is mediated by
metric perturbations.
The kinetic and mass terms
of a scalar field $\phi$
generically contain cubic interactions
schematically in the form
$\delta g_{\mu\nu} \phi^2$,
where the metric perturbations
$\delta g_{\mu\nu}$ contain
both dynamical and auxiliary fields.
These cubic interactions
induce effective quartic interactions of $\phi$
(see Fig.~\ref{fig1}),
which can be regarded as
the general relativistic counterpart
of the scalar quartic gravitational interaction in the Newtonian approximation.
We then would like to determine
the effective quartic interactions induced by gravity
and discuss their implication to axion cosmology.

For this purpose,
we first introduce a generic action
for a massive scalar field $\phi$
in the cosmological backgrounds
by the use of the effective field theory (EFT)
approach to cosmological perturbations~\cite{Cheung:2007st,Senatore:2010wk,Gubitosi:2012hu,Bloomfield:2012ff}.
In cosmology, we know that spatial homogeneity and isotropy are satisfied for our universe,
while time-translation symmetry is broken.
Using such symmetries, it is possible to constrain the structure of gravitational interactions
without knowing
the complete structure of the system.
We will see that
the obtained action contains
three types of interactions:
Two of them are cubic interactions
schematically in the forms $\gamma_{ij}\phi^2$ and $\zeta \phi^2$,
where $\gamma_{ij}$ denote the tensor degrees of freedom
(gravitational waves)
and $\zeta$ is the adiabatic mode
(i.e. the fluctuation of radiations for the radiation-dominated background).
The last one is the quartic interaction of $\phi$
obtained after solving constraints associated with
auxiliary fields.
Using these interactions,
we then evaluate the effective quartic
interactions induced by gravity
and determine the Hamiltonian
for these effective gravitational interactions.
Once we write down the interaction Hamiltonian, it is straightforward to compute the 
interaction rate
in the same manner as the previous study~\cite{Saikawa:2012uk}.
Note that the formulation in the first half of the paper is aimed at a general massive scalar field in the Friedmann-Robertson-Walker (FRW) background,
and it can also be applied to other classes of models with a subdominant scalar field, such as the dynamics of
the scalar field generating the baryon asymmetry~\cite{Affleck:1984fy},
the scalar field whose fluctuations are responsible for primordial curvature perturbations~\cite{Enqvist:2001zp}, and so on.

The outline of the paper is as follows:
In Sec.~\ref{sec2}, we first define the system with a massive scalar field (axion) in the FRW
background and construct the action, including its gravitational interactions.
Based on the action obtained there, we also derive the interaction Hamiltonian containing the cubic interactions
between the scalar and gravitons.
In Sec.~\ref{sec3}, we extract relevant interactions acting in the system of axions in the condensed regime
and derive the effective quartic interaction induced by the graviton exchange.
This formulation is applied to the system of dark matter axions to estimate their interaction rate in Sec.~\ref{sec4}.
As a result of calculations, we will show that the rate of the process caused by modes inside the horizon reproduces the previous result [Eq.~\eqref{eq1-1}]
obtained based on the Newtonian gravity.
Finally, Sec.~\ref{sec5} is devoted to summary and discussions.
Three appendixes consist of miscellaneous topics:
Some details of the tensor calculations used in Sec.~\ref{sec2} are summarized in Appendix~\ref{secA}.
In Appendix~\ref{secB}, we give expressions for mode expansions of the fields $\phi$ and $\zeta$, which are used to calculate the expectation values
in the interaction picture.
In Appendix~\ref{secC}, we discuss the interaction rate for modes outside the horizon, which is not focused on in the main text of the paper.

\section{\label{sec2} Massive scalar field
in the cosmological background}
In this section
we introduce our setup
to discuss gravitational interactions
of massive scalar fields
in the cosmological background.
By the use of the effective field theory approach
to cosmological perturbations~\cite{Cheung:2007st},
a generic action of a massive scalar field $\phi$
(and the metric $g_{\mu\nu}$)
in the FRW background
is first constructed in Sec.~\ref{sec2-1}.
In our construction,
we assume that
the effects of the scalar field
on the background dynamics are negligible
and the background evolution is determined
by some other fluid components, such as radiations.
The evolution of background fluids
breaks the time-diffeomorphism invariance
so that
one can take the unitary gauge,
in which fluctuations of fluids are absent
and they are encoded in the metric perturbations.
The propagating physical degrees of freedom
in such a unitary gauge
are then
the two transverse modes and
one longitudinal mode of the graviton,
and the massive scalar field~$\phi$.
In Sec.~\ref{sec2-2},
we write down the action
in terms of these four physical modes
by solving the momentum and the Hamiltonian
constraints.
The Hamiltonian
in the interaction picture
is then introduced in Sec.~\ref{sec2-3}.

Although we are interested in the dynamics of dark matter axions,
in the following two sections
we do not specify $\phi$ as the axion:
the results there
are applicable to general massive scalar fields
in the time-evolving background.
After developing general discussions
in these two sections,
we apply our results
to the system of coherently oscillating axions in the radiation-dominated universe in Sec.~\ref{sec4}.

\subsection{\label{sec2-1}Action from effective field theory approach}
Let us first introduce a generic action
for a massive scalar field $\phi$
in the FRW background
via the effective field theory approach
to cosmological perturbations~\cite{Cheung:2007st}.
As we mentioned above,
we assume that
the FRW background geometry
is supported by some other fluids
and the background dynamics of $\phi$
is negligible.
For the construction of the generic action,
it is convenient to take the unitary gauge,
where there are no fluid perturbations
and
the adiabatic perturbations are described
by metric perturbations.
In such a unitary gauge,
degrees of freedom
relevant to gravitational interactions of $\phi$
would be those of the metric $g_{\mu\nu}$
and the massive scalar field $\phi$.
We then write the action for these perturbations
schematically as
\begin{align}
S&=S_\phi+S_{\rm grav}\,, 
\end{align}
where $S_\phi$ contains both the metric $g_{\mu\nu}$ and the massive scalar $\phi$, and $S_{\rm grav}$ contains $g_{\mu\nu}$ only.
Assuming that the massive scalar field
is coupled to the fluids
not directly but only through the gravitational interaction,
we consider the following action for $\phi$:
\begin{align}
S_\phi&=\int d^4x\sqrt{-g}\left[-\frac{1}{2}g^{\mu\nu}\partial_\mu\phi\partial_\nu\phi
-\frac{1}{2}m^2\phi^2\right]\,,
\end{align}
where $m$ is the mass of the scalar field.
After taking the unitary gauge,
we still have the time-dependent spatial diffeomorphism invariance,
which is not broken by the fluid evolution.
The action $S_{\rm grav}$
for the metric perturbations
in the unitary gauge
would then be determined
by this residual symmetry.
As discussed in Ref.~\cite{Cheung:2007st},
it can be expanded systematically
in perturbations and derivatives.
At the lowest order in perturbations,
the action~$S_{\rm grav}$ can be determined
by the background equations of motion
as
\begin{align}
S_{\rm grav}=\int d^4x \sqrt{-g}\left[
\frac{1}{2}M_{\rm Pl}^2R
+M_{\rm Pl}^2\dot{H}g^{00}
-M_{\rm Pl}^2(3H^2+\dot{H})
\right]\,,\label{eq2-1-3}
\end{align}
where the Plank mass $M_{\rm Pl}$ is
related to Newton's constant $G$ as
$M_{\rm Pl}^2=(8\pi G)^{-1}$
and it is time independent.\footnote{
In general,
the coefficient in front of the Ricci scalar $R$
in Eq.~\eqref{eq2-1-3}
can depend on the time coordinate $t$.
However,
the time dependence can be removed
by the redefinition of the metric $g_{\mu\nu}$
and the time coordinate $t$.
See e.g. Appendix C of Ref.~\cite{Gubitosi:2012hu} for details.}
The Hubble parameter $H(t)$ is
that of the background FRW spacetime:
\begin{equation}
ds^2=-dt^2+a(t)^2dx^2
\quad
{\rm with}
\quad
H(t)=\frac{\dot{a}}{a}\,, \label{eq2-1-4}
\end{equation}
where the background scale factor $a(t)$
depends only on the time coordinate $t$.
While the three terms displayed in Eq.~\eqref{eq2-1-3}
specify the background dynamics
such as the background metric
and the background energy-momentum tensor
of the fluids,
details of perturbations
are encoded in higher-order terms.
For example,
the following term is relevant to the sound speed of adiabatic perturbations:
\begin{align}
\int d^4x \sqrt{-g}\left[
\frac{M_2(t)^4}{2}\left(g^{00}+1\right)^2
\right]\,.
\label{relevant_to_ss}
\end{align}
Here $M_2(t)$ is a time-dependent free parameter
of the theory,
and Eq.~\eqref{relevant_to_ss}
contains the second- and higher-order
metric perturbations.
Since the sound speed of
adiabatic perturbations
is not unity in the radiation-dominated universe, for example,
we take this term into account
in the following discussions.\footnote{
In addition to Eq.~\eqref{relevant_to_ss},
there are some interactions relevant to
the sound speed of adiabatic perturbations:
the interaction
$\left(g^{00}+1\right)\left(K-3H^2\right)$
can change the sound speed for example,
where $K=g^{\mu\nu}K_{\mu\nu}$ is the trace part
of the extrinsic curvature on the constant-$t$ surfaces.
However,
we do not consider such interactions in this paper
because they are higher order in derivatives.}
As we will see,
the sound speed $c_s$ of adiabatic perturbations
is determined by the parameter $M_2(t)$ as
\begin{align}
c_s^{2}&=\frac{-M_{\rm Pl}^2\dot{H}}{-M_{\rm Pl}^2\dot{H}+2M_2^4}\,.
\end{align}
By further assuming that
tensor perturbations have canonical dispersion relations,
we drop other second-order terms
in perturbations.
The third- and higher-order metric perturbations
are not relevant for our purpose,
because we would like to determine
effective scalar quartic interaction
originated from cubic interactions
in the form $\delta g_{\mu\nu}\phi^2$.
We therefore employ the following action
to discuss gravitational interactions
of the massive scalar field $\phi$:
\begin{align}
S&=S_\phi+S_{\rm grav}\,,
\label{eq2-1-1}\\
S_\phi&=\int d^4x\sqrt{-g}\left[-\frac{1}{2}g^{\mu\nu}\partial_\mu\phi\partial_\nu\phi
-\frac{1}{2}m^2\phi^2\right]\,,
\label{eq2-1-2}\\
S_{\rm grav}&=\int d^4x \sqrt{-g}\left[
\frac{1}{2}M_{\rm Pl}^2R
+M_{\rm Pl}^2\dot{H}g^{00}
-M_{\rm Pl}^2(3H^2+\dot{H})
+\frac{M_2(t)^4}{2}\left(g^{00}+1\right)^2
\right]\,.\label{eq2-1-5}
\end{align}

We then rewrite the above action
in terms of the Arnowitt-Deser-Misner (ADM) decomposition~\cite{Arnowitt:1962hi}:
\begin{equation}
ds^2=-(N^2-N_iN^i)dt^2+2N_i dx^i dt+h_{ij}\,dx^idx^j
\,. \label{eq2-1-6}
\end{equation}
Here and in what follows we use the spatial metric $h_{ij}$ and its inverse $h^{ij}$ to
raise or lower the indices of $N^i$.
The inverse metric $g^{\mu\nu}$
is expressed in terms of $N$, $N^i$, and $h_{ij}$ as
\begin{equation}
g^{00}=-\frac{1}{N^2}\,,
\quad
g^{0i}=g^{i0}=\frac{N^i}{N^2}\,,
\quad
g^{ij}=h^{ij}-\frac{N^iN^j}{N^2}\,. \label{eq2-1-7}
\end{equation}
The four-dimensional Ricci scalar $R$ can be written as
\begin{align}
R&=R^{(3)}+N^{-2}(E_{ij}E^{ij}-E^2)+(\text{total derivatives})\,, \label{eq2-1-8}
\end{align}
where $R^{(3)}$ is the three-dimensional Ricci scalar.
The quantities $E_{ij}$ and $E$ are defined by
\begin{align}
E_{ij}&=\frac{1}{2}\left(\dot{h}_{ij}-\nabla^{(3)}_i N_j-\nabla^{(3)}_j N_i\right)\,,
\quad
E = E^i_i\,,\label{eq2-1-9}
\end{align}
where $\nabla^{(3)}$ is the three-dimensional covariant derivative
and $E_{ij}$
is related to the extrinsic curvature $K_{ij}$ as $E_{ij}=NK_{ij}$.
Using the decomposition [Eq.~\eqref{eq2-1-6}], we rewrite the action as
\begin{align}
S_\phi&=\int d^4x\,\sqrt{h}\Bigg[
\frac{1}{2}N^{-1}\dot{\phi}^2
-\frac{N^i}{N}\dot{\phi}\partial_i\phi
-\frac{1}{2}\Big(N\,h^{ij}-\frac{N^iN^j}{N}\Big)\partial_i\phi\partial_j\phi
-\frac{1}{2}N\,m^2\phi^2\Bigg]\,,\label{eq2-1-10}\\
S_{\rm grav}&=M_{\rm Pl}^2\int d^4x \sqrt{h}\left[
\frac{1}{2}NR^{(3)}
+\frac{1}{2}N^{-1}(E_{ij}E^{ij}-E^2)-N^{-1}\dot{H}
-N(3H^2+\dot{H})
+\frac{M_2^4}{2M_{\rm Pl}^2}\left(N^{-2}-1\right)^2
\right]\,.
\label{eq2-1-11}
\end{align}
To fix the residual gauge symmetry
associated with the time-dependent spatial diffeomorphism,
let us impose the following transverse conditions:
\begin{equation}
h_{ij}=a^2e^{2\zeta} (e^\gamma)_{ij}
\quad
{\rm with}
\quad
\gamma_{ii}=\partial_i\gamma_{ij}=0.
\end{equation}
Then,
the action contains
four propagating physical modes:
the massive scalar $\phi$,
the adiabatic mode $\zeta$,
and
the two tensor modes $\gamma_{ij}$.
In addition to these four physical modes,
we have auxiliary fields $N$ and $N^i$,
which do not have kinetic terms.
In the next subsection,
we solve the Hamiltonian and the momentum constraints
associated with these auxiliary fields
and rewrite the action
in terms of $\phi$, $\zeta$, and $\gamma$.

\subsection{\label{sec2-2} Action for dynamical fields}
Since the action in Eq.~\eqref{eq2-1-11} does not contain the kinetic terms for $N$ and $N^i$,
they are regarded as auxiliary fields.
We then solve the constraints
associated with these auxiliary fields:
\begin{align}
\frac{\delta S}{\delta N}=
\frac{\delta S}{\delta N^i}=0. \label{eq2-2-1}
\end{align}
For our purpose,
we need the quadratic action of $\phi$,
that of metric perturbations $\delta g_{\mu\nu}$,
and the cubic interaction in the form $\delta g_{\mu\nu}\phi^2$.
We therefore expand the action up to this order
and solve the constraints.
Let us first rewrite $N$ and $N^i$ as
\begin{align}
N=1+N_1\,,
\quad
N^i=\partial^i\psi+N_T^i
\quad
{\rm with}
\quad
\partial_iN_T^i=0,
\end{align}
where $N_1$, $N^i_T$, and $\psi$
are of the first order in perturbations.
Using these variables,
the action [Eqs.~\eqref{eq2-1-10},~\eqref{eq2-1-11}]
can be expanded as follows
(see Appendix~\ref{secA} for the details of the calculation):
\begin{align}
\nonumber
S_{\rm grav}&=
M_{\rm Pl}^2\int d^4x\,a^3\Bigg[
-3\dot{\zeta}^2+\frac{(\partial_i\zeta)^2}{a^2}
+6HN_1\dot{\zeta}
-2N_1\frac{\partial^2\zeta}{a^2}
-\Big(3H^2
+\dot{H}-\frac{2M_2^4}{M_{\rm Pl}^2}\Big)N_1^2\\
&\qquad\qquad\qquad\qquad
+\Big(2\dot{\zeta}-2HN_1\Big)\frac{\partial^2\psi}{a^2}
+\frac{1}{4}(\partial_jN_T^i)(\partial_jN_T^i)
+\frac{1}{8}\Big(\dot{\gamma}_{ij}^2-\frac{(\partial_k\gamma_{ij})^2}{a^2}\Big)
\Bigg]\,,
\label{deltag_second}
\\
\nonumber
S_\phi&=\int d^4x\,a^3\Bigg[
\frac{1}{2}\dot{\phi}^2
-\frac{1}{2}\frac{(\partial_i\phi)^2}{a^2}
-\frac{1}{2}m^2\phi^2
-\frac{1}{2}N_1\Big(\dot{\phi}^2+\frac{(\partial_i\phi)^2}{a^2}+m^2\phi^2\Big)
\\
\label{phi_cubic}
&\qquad\qquad\qquad
+\frac{1}{2}\zeta \Big(3\dot{\phi}^2-\frac{(\partial_i\phi)^2}{a^2}
-3m^2 \phi^2\Big)
+\frac{1}{2}\gamma_{ij}\frac{\partial_i\phi\partial_j\phi}{a^2}
-\frac{\partial_i\psi}{a^2}\dot{\phi}\partial_i\phi
+N^i_T\frac{\partial_j}{\partial^{2}}(\partial_i\dot{\phi}\partial_j\phi
-\partial_j\dot{\phi}\partial_i\phi)
\Bigg]\,,
\end{align}
where
$\partial^2=\partial_i^2$
and
we have dropped
temporal and spatial total derivatives.
To obtain the expression~\eqref{phi_cubic},
we divided $\dot{\phi}\partial_i\phi$
into the transverse part and the $\partial_i$-exact part
as
\begin{align}
\dot{\phi}\partial_i\phi
=-\frac{\partial_j}{\partial^{2}}(\partial_i\dot{\phi}\partial_j\phi
-\partial_j\dot{\phi}\partial_i\phi)
+\frac{\partial_i\partial_j}{\partial^{2}}(\dot{\phi}\partial_j\phi),
\end{align}
and used
\begin{align}
\int d^4x\,a^3N_T^i\dot{\phi}\partial_i\phi
=-\int d^4x\,a^3N_T^i\frac{\partial_j}{\partial^{2}}(\partial_i\dot{\phi}\partial_j\phi
-\partial_j\dot{\phi}\partial_i\phi).
\end{align}
Then,
variations of the action [Eqs.~\eqref{deltag_second},~\eqref{phi_cubic}]
with respect to auxiliary fields $N_1$, $N_T^i$,
and $\psi$
lead to the following constraints:
\begin{align}
M_{\rm Pl}^2\left[
6H\dot{\zeta}
-2\frac{\partial^2\zeta}{a^2}
-\Big(6H^2+2\dot{H}
-\frac{4M_2^4}{M_{\rm Pl}^2}\Big)N_1
-2H\frac{\partial^2\psi}{a^2}
\right]
-\frac{1}{2}\Big(\dot{\phi}^2
+\frac{(\partial_i\phi)^2}{a^2}
+m^2\phi^2\Big)&=0,
\\
\frac{M_{\rm Pl}^2}{2}\partial^2N^i_T
-\frac{\partial_j}{\partial^{2}}(\partial_i\dot{\phi}\partial_j\phi
-\partial_j\dot{\phi}\partial_i\phi)
&=0,\\
M_{\rm Pl}^2\left[-2\partial^2\dot{\zeta}
+2H\partial^2N_1
\right]
-\partial_i(\dot{\phi}\partial_i\phi)
&=0.
\end{align}
Solving these constraints,
we find
\begin{equation}
N_1=\frac{\dot{\zeta}}{H}
+\frac{1}{2M_{\rm Pl}^2H}
\frac{\partial_i}{\partial^{2}}(\dot{\phi}\partial_i\phi)\,,
\quad
N^i_T=\frac{2}{M_{\rm Pl}^2}\frac{\partial_j}{(\partial^{2})^2}(\partial_i\dot{\phi}\partial_j\phi-\partial_j\dot{\phi}\partial_i\phi)\,,
\quad
\frac{\psi}{a^2}=-a^{-2}\frac{\zeta}{H}+\chi\,, \label{eq2-2-10}
\end{equation}
where $\chi$ is defined by
\begin{align}
\partial^2\chi=-\frac{\dot{H}}{H^2}\dot{\zeta}
+\frac{2M_2^4}{M_{\rm Pl}^2H^2}\dot{\zeta}
-\frac{1}{4M_{\rm Pl}^2H}\left[
\dot{\phi}^2
+\frac{(\partial_i\phi)^2}{a^2}
+m^2\phi^2
+\Big(6H+2\frac{\dot{H}}{H}
-\frac{4M_2^4}{M_{\rm Pl}^2H}\Big)
\partial^{-2}\partial_i
(\dot{\phi}\partial_i\phi)
\right]\,. \label{eq2-2-11}
\end{align}
Substituting the constraints~\eqref{eq2-2-10} and~\eqref{eq2-2-11} into
the action [Eqs.~\eqref{deltag_second},~\eqref{phi_cubic}],
we obtain
\begin{align}
\nonumber
S&=S_{\phi}+S_{\rm grav}\\
&=S_{\rm free}+S_{\zeta\phi^2}+S_{\gamma\phi^2}+S_{\phi^4}\,, \label{eq2-3-12}
\\
S_{\rm free}&=\int d^4x \,a^3\left[
M_{\rm Pl}^2\tilde{\epsilon}\Big(\dot{\zeta}^2-c_s^2\frac{(\partial_i\zeta)^2}{a^2}\Big)
+\frac{M_{\rm Pl}^2}{8}\Big(\dot{\gamma}_{ij}^2-\frac{(\partial_k\gamma_{ij})^2}{a^2}\Big)
+\frac{1}{2}\Big(\dot{\phi}^2-\frac{(\partial_i\phi)^2}{a^2}-m^2\phi^2\Big)
\right]\,, \label{eq2-3-14}
\\
S_{\zeta\phi^2}&=\int d^4x\,a^3\left[
\frac{1}{2}\zeta\Big(3\dot{\phi}^2-\frac{(\partial_i\phi)^2}{a^2}-3m^2\phi^2\Big)
-\frac{1}{2H}\dot{\zeta}\Big(\dot{\phi}^2+\frac{(\partial_i\phi)^2}{a^2}+m^2\phi^2\Big)
+\Big(\tilde{\epsilon}\dot{\zeta}-\frac{1}{H}\frac{\partial^2\zeta}{a^2}\Big)
\Big(\partial^{-2}\partial_i(\dot{\phi}\partial_i\phi)\Big)
\right]\,, \label{eq2-3-18}
\\
S_{\gamma\phi^2}&=\int d^4x\,a^3
\left[\frac{1}{2}\gamma_{ij}\frac{\partial_i\phi\partial_j\phi}{a^2}\right]\,, \label{eq2-3-16}
\\
S_{\phi^4}&=\int d^4x \,a^3\left[
\frac{1+\tilde{\epsilon}}{4M_{\rm Pl}^2}
\Big(\partial^{-2}\partial_i(\dot{\phi}\partial_i\phi)\Big)^2
+\frac{1}{M_{\rm Pl}^2}(\dot{\phi}\partial_i\phi)\partial^{-2}(\dot{\phi}\partial_i\phi)
-\frac{1}{4M_{\rm Pl}^2H}\Big(\dot{\phi}^2+\frac{(\partial_i\phi)^2}{a^2}+m^2\phi^2\Big)
\partial^{-2}\partial_i(\dot{\phi}\partial_i\phi)
\right]\,,\label{eq2-3-19}
\end{align}
where we have introduced
the sound speed~$c_s$
of adiabatic perturbations
and the parameter $\tilde{\epsilon}$
in analogy with the slow-roll parameter
$\epsilon=-\dot{H}/H^2$
as
\begin{align}
c_s^2=\frac{-M_{\rm Pl}^2\dot{H}}{-M_{\rm Pl}^2\dot{H}+2M_2^4}\,,
\quad
\tilde{\epsilon}=c_s^{-2}\epsilon
=-c_s^{-2}\frac{\dot{H}}{H^2}\,.
\end{align}
As we mentioned in the Introduction,
the action in Eq.~\eqref{eq2-3-12} contains
three types of interactions:
cubic interactions
in the forms
$\zeta \phi^2$ and $\gamma_{ij}\phi^2$,
and the scalar quartic interaction.
Note that when $M_2(t)=0$,
the sound speed becomes unity $c_s=1$
and the parameter $\tilde{\epsilon}$
reduces to the usual slow-roll parameter $\tilde{\epsilon}=\epsilon$.
On the other hand, in Sec.~\ref{sec4} we will consider the system where the background fluid is dominated by radiations.
In such a case we must take the value of the sound speed as $c_s\simeq 1/\sqrt{3}$.

\subsection{\label{sec2-3}Hamiltonian in the interaction picture}
We then introduce
the Hamiltonian in the interaction picture
for our system.
Since the cubic and quartic terms
($S_{\zeta\phi^2}$, $S_{\gamma\phi^2}$,
and $S_{\phi^4}$)
contain derivative couplings,
the interaction Hamiltonian $H_{\rm int}$
does not coincide with
the negative of the interaction terms in the Lagrangian,
$-L_{\rm int}$.
Although the construction
of the interaction Hamiltonian
can be performed straightforwardly
as in standard textbooks,
we first introduce a
useful formula to
calculate the interaction Hamiltonian
for the system with derivative couplings,
and then we apply it to our system.

\subsubsection{A useful formula}
Let us begin by considering the following action
\begin{align}
S=\int d^4x \mathcal{L}(\phi_a,\dot{\phi}_a)
=\int d^4x \left[\mathcal{L}_{\rm free}(\phi_a,\dot{\phi}_a)
+\mathcal{L}_{\rm int}(\phi_a,\dot{\phi}_a)\right]\,, \label{eq3-1}
\end{align}
where the index $a$ stands for the fields involved in the model (such as $\phi$, $\zeta$, and $\gamma$) and their spin components.
We suppose that
the free Lagrangian $\mathcal{L}_{\rm free}$
is constructed from the temporal kinetic terms $\displaystyle\sum_{a}\frac{\alpha_a^2}{2}\dot{\phi}_a^2$
and the part $\mathcal{L}_{\rm free,\phi}(\phi_a)$
without $\dot{\phi}_a$ as
\begin{align}
\mathcal{L}_{\rm free}&=\sum_{a}\frac{\alpha_a^2}{2}\dot{\phi}_a^2
+\mathcal{L}_{\rm free,\phi}(\phi_a)\,.
\end{align}
We also assume that
the interaction Lagrangian $\mathcal{L}_{\rm int}(\phi_a,\dot{\phi}_a)$
does not contain terms with more than second-order time derivatives.
Then, the momenta $\pi_a$'s are given by
\begin{align}
\pi_a&=\frac{\partial\mathcal{L}}{\partial\dot{\phi}_a}(\phi_a,\dot{\phi}_a)
=\alpha_a^2\,\dot{\phi}_a+\frac{\partial\mathcal{L}_{\rm int}}{\partial\dot{\phi}_a}(\phi_a,\dot{\phi}_a)\,. \label{eq3-2}
\end{align}
The Hamiltonian density $\mathcal{H}$ is written as
\begin{align}
\mathcal{H}&=\sum_a\pi_a\dot{\phi}_a-\mathcal{L}(\phi_a,\dot{\phi}_a)\nonumber \\
&=\sum_a\alpha_a^{-2}\pi_a\Big(\pi_a-\frac{\partial\mathcal{L}_{\rm int}}{\partial\dot{\phi}_a}(\phi_a,\dot{\phi}_a)\Big)
-\sum_a\frac{\alpha_a^{-2}}{2}\Big(\pi_a-\frac{\partial\mathcal{L}_{\rm int}}{\partial\dot{\phi}_a}(\phi_a,\dot{\phi}_a)\Big)^2
-\mathcal{L}_{\rm free,\phi}(\phi_a)-\mathcal{L}_{\rm int}(\phi_a,\dot{\phi}_a)\nonumber \\
&=\sum_a\frac{\alpha_a^{-2}}{2}\pi_a^2
-\mathcal{L}_{\rm free,\phi}(\phi_a)
-\mathcal{L}_{\rm int}(\phi_a,\dot{\phi}_a)
-\sum_a\frac{\alpha_a^{-2}}{2}\left(\frac{\partial\mathcal{L}_{\rm int}}{\partial\dot{\phi}_a}(\phi_a,\dot{\phi}_a)\right)^2\,. \label{eq3-3}
\end{align}
Let us Taylor-expand as follows:
\begin{align}
\nonumber
\mathcal{L}_{\rm int}(\phi_a,\dot{\phi}_a)
&=
\mathcal{L}_{\rm int}
\left(\phi_a,\,\alpha^{-2}_a\pi_a
-\alpha^{-2}_a\frac{\partial\mathcal{L}_{\rm int}}{\partial\dot{\phi}_a}(\phi_a,\dot{\phi}_a)\right)\\
\nonumber
&=\mathcal{L}_{\rm int}
\left(\phi_a,\,\alpha^{-2}_a\pi_a\right)
-\sum_a\alpha_a^{-2}
\frac{\partial\mathcal{L}_{\rm int}}{\partial\dot{\phi}_a}(\phi_a,\alpha_a^{-2}\pi_a)\,
\frac{\partial\mathcal{L}_{\rm int}}{\partial\dot{\phi}_a}(\phi_a,\dot{\phi}_a)\,\\
&\quad
+\frac{1}{2}\sum_{a,b}\alpha_a^{-2}\alpha_b^{-2}\frac{\partial^2\mathcal{L}_{\rm int}}{\partial\dot{\phi}_a\partial\dot{\phi}_b}(\phi_a,\alpha_a^{-2}\pi_a)\,\frac{\partial\mathcal{L}_{\rm int}}{\partial\dot{\phi}_a}(\phi_a,\dot{\phi}_a)\,
\frac{\partial\mathcal{L}_{\rm int}}{\partial\dot{\phi}_b}(\phi_a,\dot{\phi}_a)\,, \label{eq3-4}\\
\nonumber
\frac{\partial\mathcal{L}_{\rm int}}{\partial\dot{\phi}_a}(\phi_a,\dot{\phi}_a)
&=
\frac{\partial\mathcal{L}_{\rm int}}{\partial\dot{\phi}_a}\left(\phi_a,\,\alpha^{-2}_a\pi_a-\alpha^{-2}_a\frac{\partial\mathcal{L}_{\rm int}}{\partial\dot{\phi}_a}(\phi_a,\dot{\phi}_a)\right)\\
&=\frac{\partial\mathcal{L}_{\rm int}}{\partial\dot{\phi}_a}\left(\phi_a,\,\alpha^{-2}_a\pi_a\right)-\sum_b\alpha_b^{-2}\frac{\partial^2\mathcal{L}_{\rm int}}{\partial\dot{\phi}_a\partial\dot{\phi}_b}\left(\phi_a,\,\alpha^{-2}_a\pi_a\right)\frac{\partial\mathcal{L}_{\rm int}}{\partial\dot{\phi}_b}(\phi_a,\dot{\phi}_a)\,. \label{eq3-5}
\end{align}
We then have
\begin{align}
\mathcal{H}&=\sum_a\frac{\alpha_a^{-2}}{2}\pi_a^2
-\mathcal{L}_{\rm free,\phi}(\phi_a)
-\mathcal{L}_{\rm int}(\phi_a,\alpha_a^{-2}\pi_a)
+\frac{1}{2}\sum_a\alpha_a^{-2}
\left(\frac{\partial\mathcal{L}_{\rm int}}{\partial\dot{\phi}_a}(\phi_a,\alpha_a^{-2}\pi_a)\right)
\left(\frac{\partial\mathcal{L}_{\rm int}}{\partial\dot{\phi}_a}(\phi_a,\dot{\phi}_a)\right)\,. 
\end{align}
It is also possible to expand the last term
as follows:
\begin{align}
\nonumber
\mathcal{H}&=\sum_a\frac{\alpha_a^{-2}}{2}\pi_a^2
-\mathcal{L}_{\rm free,\phi}(\phi_a)
-\mathcal{L}_{\rm int}(\phi_a,\alpha_a^{-2}\pi_a)
+\frac{1}{2}\sum_a\alpha_a^{-2}
\frac{\partial\mathcal{L}_{\rm int}}{\partial\dot{\phi}_a}(\phi_a,\alpha_a^{-2}\pi_a)\,
\frac{\partial\mathcal{L}_{\rm int}}{\partial\dot{\phi}_a}(\phi_a,\alpha_a^{-2}\pi_a)\\
\nonumber
&\quad
-\frac{1}{2}\sum_{a,b}\alpha_a^{-2}\alpha_b^{-2}
\frac{\partial\mathcal{L}_{\rm int}}{\partial\dot{\phi}_a}(\phi_a,\alpha_a^{-2}\pi_a)\,
\frac{\partial\mathcal{L}_{\rm int}}{\partial\dot{\phi}_a\partial\dot{\phi}_b}(\phi_a,\alpha_a^{-2}\pi_a)\,
\frac{\partial\mathcal{L}_{\rm int}}{\partial\dot{\phi}_b}(\phi_a,\alpha_a^{-2}\pi_a)\\
&\quad
+\frac{1}{2}\sum_{a,b,c}\alpha_a^{-2}\alpha_b^{-2}\alpha_c^{-2}
\frac{\partial\mathcal{L}_{\rm int}}{\partial\dot{\phi}_a}(\phi_a,\alpha_a^{-2}\pi_a)\,
\frac{\partial\mathcal{L}_{\rm int}}{\partial\dot{\phi}_a\partial\dot{\phi}_b}(\phi_a,\alpha_a^{-2}\pi_a)\,
\frac{\partial\mathcal{L}_{\rm int}}{\partial\dot{\phi}_b\partial\dot{\phi}_c}(\phi_a,\alpha_a^{-2}\pi_a)\,
\frac{\partial\mathcal{L}_{\rm int}}{\partial\dot{\phi}_c}(\phi_a,\alpha_a^{-2}\pi_a)+\ldots\,. \label{eq3-1-6}
\end{align}

Now let us take the free Hamiltonian $\mathcal{H}_{\rm free}$
and the interaction Hamiltonian $\mathcal{H}_{\rm int}$
as
\begin{align}
\mathcal{H}_{\rm free}&=\sum_a\frac{\alpha_a^{-2}}{2}\pi_a^2
-\mathcal{L}_{\rm free,\phi}(\phi_a)\,,\\
\mathcal{H}_{\rm int}&=-\mathcal{L}_{\rm int}(\phi_a,\alpha_a^{-2}\pi_a)
+\frac{1}{2}\sum_a\alpha_a^{-2}
\left(\frac{\partial\mathcal{L}_{\rm int}}{\partial\dot{\phi}_a}(\phi_a,\alpha_a^{-2}\pi_a)\right)
\left(\frac{\partial\mathcal{L}_{\rm int}}{\partial\dot{\phi}_a}(\phi_a,\dot{\phi}_a)\right)\,.
\end{align}
Then,
the canonical momenta $\pi_a^I$'s
in the interaction picture
are given by
\begin{align}
\pi_a^I=\alpha_a^2\dot{\phi}_a^I\,,
\end{align}
where $\phi_a^I$'s are the canonical fields
in the interaction picture
and their evolution is determined by
the free field equations of motion.
In terms of $\phi_a^I$'s,
the interaction Hamiltonian can be written as
\begin{align}
\nonumber
\mathcal{H}_{\rm int}&=-\mathcal{L}_{\rm int}(\phi_a^I,\dot\phi_a^I)
+\frac{1}{2}\sum_a\alpha_a^{-2}
\left(\frac{\partial\mathcal{L}_{\rm int}}{\partial\dot{\phi}_a}(\phi_a^I,\dot\phi_a^I)\right)
\left(\frac{\partial\mathcal{L}_{\rm int}}{\partial\dot{\phi}_a}(\phi_a^I,\dot\phi_a^I)\right)\\
&\quad
-\frac{1}{2}\sum_{a,b}\alpha_a^{-2}\alpha_b^{-2}
\frac{\partial\mathcal{L}_{\rm int}}{\partial\dot{\phi}_a}(\phi_a^I,\dot\phi_a^I)\,
\frac{\partial\mathcal{L}_{\rm int}}{\partial\dot{\phi}_a\partial\dot{\phi}_b}(\phi_a^I,\dot\phi_a^I)\,
\frac{\partial\mathcal{L}_{\rm int}}{\partial\dot{\phi}_b}(\phi_a^I,\dot\phi_a^I)+\ldots\,,
\label{eq3-6}
\end{align}
where we have used the expansion in Eq.~\eqref{eq3-1-6}.

\subsubsection{Application to the scalar graviton system}
Let us next construct the interaction Hamiltonian
for our system
by taking the free part as in Eq.~\eqref{eq2-3-14}
and
using the formula in Eq.~\eqref{eq3-6}.
For our purpose,
we need to determine the interaction Hamiltonian
in the forms $\zeta\phi^2$, $\gamma\phi^2$, and $\phi^4$.
Since our action contains cubic and higher interactions only,
the second line of Eq.~\eqref{eq3-6}
becomes the quintic and higher order.
Therefore,
the interaction Hamiltonian density
up to the quartic order
is given by
\begin{align}
\mathcal{H}_{\rm int}&=
-\mathcal{L}_{\rm int}(\phi_a^I,\dot\phi_a^I)
+\frac{1}{2}\sum_a\alpha_a^{-2}
\left(\frac{\partial\mathcal{L}_{\rm int}^{(3)}}{\partial\dot{\phi}_a}(\phi_a^I,\dot\phi_a^I)\right)
\left(\frac{\partial\mathcal{L}_{\rm int}^{(3)}}{\partial\dot{\phi}_a}(\phi_a^I,\dot\phi_a^I)\right)\,. \label{eq3-8}
\end{align}
We also notice that
the time derivative of the scalar field~$\dot{\phi}$
in the interactions
always appears with $\zeta$,
so that the $\dot{\phi}$ contractions
in the last term of Eq.~\eqref{eq3-8}
lead to interaction terms with at least two $\zeta$'s.
Since such interactions are irrelevant to our discussions,
we can neglect~$\dot{\phi}$ contractions,
and the only contribution we have to care about is
that from the $\dot{\zeta}$ contractions.
It is straightforward to calculate this contribution:
\begin{align}
\nonumber
&\frac{1}{2}\alpha_\zeta^{-2}
\left(\frac{\partial\mathcal{L}_{\zeta\phi^2}}{\partial\dot{\zeta}}(\phi_a^I,\dot\phi_a^I)\right)
\left(\frac{\partial\mathcal{L}_{\zeta\phi^2}}{\partial\dot{\zeta}}(\phi_a^I,\dot\phi_a^I)\right)\\
\nonumber
&=
\frac{1}{2}(2a^3M_{\rm Pl}^2\tilde{\epsilon})^{-1}
a^6\left[
-\frac{1}{2H}\Big(\dot{\phi}^2+\frac{(\partial_i\phi)^2}{a^2}+m^2\phi^2\Big)
+\tilde{\epsilon}
\Big(\partial^{-2}\partial_i(\dot{\phi}\partial_i\phi)\Big)\right]^2
\\
\nonumber
&=a^3\left[
\frac{\tilde{\epsilon}}{4M_{\rm Pl}^2}
\Big(\partial^{-2}\partial_i(\dot{\phi}\partial_i\phi)\Big)^2
-\frac{1}{4M_{\rm Pl}^2H}
\Big(\dot{\phi}^2+\frac{(\partial_i\phi)^2}{a^2}+m^2\phi^2\Big)
\Big(\partial^{-2}\partial_i(\dot{\phi}\partial_i\phi)\Big)
\right.\\*
&\qquad\qquad\quad
\left.
+\frac{1}{16M_{\rm Pl}^2H^2\tilde{\epsilon}}
\Big(\dot{\phi}^2+\frac{(\partial_i\phi)^2}{a^2}+m^2\phi^2\Big)^2
\right]\,.
\label{eq3-10}
\end{align}
Here and in what follows,
we drop the superscript $I$
indicating that fields are in the interaction picture
for simplicity.
We therefore obtain the following Hamiltonian density
in the interaction picture:
\begin{align}
\mathcal{H} &= \mathcal{H}_{\rm free} + \mathcal{H}_{\rm int}\,, \label{eq3-11}\\
\mathcal{H}_{\rm free}&=
a^3\Big[
M_{\rm Pl}^2\tilde{\epsilon}\Big(\dot{\zeta}^2+c_s^2\frac{(\partial_i\zeta)^2}{a^2}\Big)
+\frac{M_{\rm Pl}^2}{8}\Big(\dot{\gamma}_{ij}^2+\frac{(\partial_k\gamma_{ij})^2}{a^2}\Big)
+\frac{1}{2}\Big(\dot{\phi}^2+\frac{(\partial_i\phi)^2}{a^2}
+m^2\phi^2\Big)\Big]\,, \label{eq3-12}\\
\mathcal{H}_{\rm int}
&=\mathcal{H}_{\zeta\phi^2}+\mathcal{H}_{\gamma\phi^2}+\mathcal{H}_{\phi^4}\,, \label{eq3-13}\\
\mathcal{H}_{\zeta\phi^2}&=
a^3\left[
-\frac{1}{2}\zeta\Big(3\dot{\phi}^2-\frac{(\partial_i\phi)^2}{a^2}-3m^2\phi^2\Big)
+\frac{1}{2H}\dot{\zeta}\Big(\dot{\phi}^2+\frac{(\partial_i\phi)^2}{a^2}+m^2\phi^2\Big)
-\Big(\tilde{\epsilon}\dot{\zeta}-\frac{1}{H}\frac{\partial^2\zeta}{a^2}\Big)
\Big(\partial^{-2}\partial_i(\dot{\phi}\partial_i\phi)\Big) \right]\,,\label{eq3-14}\\
\mathcal{H}_{\gamma\phi^2}&=
-\frac{a^3}{2}\gamma_{ij}\frac{\partial_i\phi\partial_j\phi}{a^2}\,, \label{eq3-15}\\
\mathcal{H}_{\phi^4}&=
a^3\left[
-\frac{1}{4M_{\rm Pl}^2}
\Big(\partial^{-2}\partial_i(\dot{\phi}\partial_i\phi)\Big)^2
-\frac{1}{M_{\rm Pl}^2}(\dot{\phi}\partial_i\phi)\partial^{-2}(\dot{\phi}\partial_i\phi)
+\frac{1}{16M_{\rm Pl}^2H^2\tilde{\epsilon}}
\Big(\dot{\phi}^2+\frac{(\partial_i\phi)^2}{a^2}+m^2\phi^2\Big)^2 \right]\,. \label{eq3-16}
\end{align}
As explained in Appendix~\ref{secB},
the fields $\phi$ and $\zeta$
in the interaction picture
are expanded as
\begin{align}
\phi({\bf x},t)
&=\int \frac{d^3 k}{(2\pi)^3}e^{-i{\bf k\cdot x}}\phi_{\bf k}(t)
=\int \frac{d^3 k}{(2\pi)^3}\left[e^{i{\bf k\cdot x}}\varphi_k(t)a_{\bf k} + e^{-i{\bf k\cdot x}}\varphi^*_k(t)a^{\dagger}_{\bf k} \right]\,,\\
\zeta({\bf x},t) &
=\int \frac{d^3 k}{(2\pi)^3}e^{-i{\bf k\cdot x}}\zeta_{\bf k}(t)
= \int \frac{d^3 k}{(2\pi)^3}\left[e^{i{\bf k\cdot x}}{\cal Z}_k(t)a_{z{\bf k}} + e^{-i{\bf k\cdot x}}{\cal Z}^*_k(t)a^{\dagger}_{z{\bf k}} \right]\,,
\end{align}
where $a_{\bf k}$ and $a_{z{\bf k}}$ are the annihilation operators satisfying the commutation relations
\begin{align}
&[a_{\bf k}, a^{\dagger}_{\bf k'}] = (2\pi)^3\delta^{(3)}({\bf k-k'})\,,
\quad
[a_{\bf k}, a_{\bf k'}] = [a^{\dagger}_{\bf k}, a^{\dagger}_{\bf k'}] = 0\,, \label{eq2-commak}\\
&[a_{z{\bf k}}, a^{\dagger}_{z{\bf k'}}] = (2\pi)^3\delta^{(3)}({\bf k-k'})\,,
\quad
[a_{z{\bf k}}, a_{z{\bf k'}}] = [a^{\dagger}_{z{\bf k}}, a^{\dagger}_{z{\bf k'}}] = 0\,, \label{eq2-commazk}\\
&(\mathrm{otherwise}) = 0. \label{eq2-commother}
\end{align}
The mode functions
$\varphi_k$ and ${\cal Z}_k$
satisfy the free equations of motion
and their concrete forms
are given in Appendix~\ref{secB}.

\section{Effective quartic interaction from graviton exchange}
\label{sec3}
In this section
we evaluate the effective quartic interaction
induced by the graviton exchange.
In the context of axion cosmology,
our interests are in the gravitational interaction
in the 
regime $m\gg k/a,H$,
since axions are produced when $m\gtrsim H$ is satisfied.
We therefore first discuss relevant interactions
in this regime,
and then evaluate the effective interaction
induced by these interactions.

\subsection{Relevant interactions
in the 
regime $m\gg H,k/a$}
\label{sec3-1}
Let us first discuss
which
terms in
the interaction Hamiltonian [Eqs.~\eqref{eq3-13}-\eqref{eq3-16}]
are relevant in the regime $m\gg k/a,H$.
In this regime,
the mode function $\varphi_k$ of the scalar field
is given by
\begin{align}
\label{mode_heavy_scalar}
\varphi_k\simeq\frac{1}{\sqrt{2ma^3}}e^{-imt}\,,
\end{align}
where we have dropped subleading terms
suppressed by the factor
$\displaystyle\frac{H}{m}$ or $\displaystyle\frac{k/a}{m}$.
As is implied from this expression,
spatial derivatives of $\phi$
are negligible compared with
$\dot{\phi}$ and $m\phi$,
so that leading contributions
in $H_{{\rm int},\zeta\phi^2}$ are given by
\begin{equation}
H_{{\rm int},\zeta\phi^2} \simeq \int d^3x a^3 \Bigg[ \frac{1}{2H}\dot{\zeta}\Big(\dot{\phi}^2 + m^2\phi^2\Big) - \frac{3}{2}\zeta\Big(\dot{\phi}^2-m^2\phi^2\Big) \Bigg]\,, \label{eq4-7}
\end{equation}
whose magnitudes can be estimated as ${\cal O}(m^2\zeta\phi^2)$
since $\dot{\zeta}\sim H\zeta$ and $\partial^2\zeta/a^2\sim H^2\zeta$.
On the other hand, the magnitude of the $\gamma\phi^2$-type interaction term [Eq.~\eqref{eq3-15}] is estimated as ${\cal O}(H^2\gamma\phi^2)$,
which is suppressed by a factor of ${\cal O}(H^2/m^2)$ compared with leading terms in Eq.~\eqref{eq4-7}.
We then drop this $\gamma\phi^2$-type interaction [Eq.~\eqref{eq3-15}].
Similarly,
the leading contribution in the $\phi^4$-type interaction [Eq.~\eqref{eq3-16}] is estimated as ${\cal O}(m^4\phi^4/M_{\rm Pl}^2H^2)$, which gives
\begin{equation}
H_{{\rm int},\phi^4} \simeq \int d^3x\,
a^3\frac{1}{16M_{\rm Pl}^2H^2\tilde{\epsilon}}
(\dot{\phi}^2+m^2\phi^2)^2\,. \label{eq4-8}
\end{equation}

Next,
we rewrite the leading interactions in Eqs.~\eqref{eq4-7} and~\eqref{eq4-8}
in terms of mode functions and creation/annihilation
operators,
and discuss relevant processes
induced by these leading interactions.
Let us start from the first term in Eq.~\eqref{eq4-7}:
\begin{align}
\nonumber
&\frac{a^3 }{2H}\int d^3x  \,\dot{\zeta}\Big(\dot{\phi}^2 + m^2\phi^2\Big)\\
\nonumber
&=\frac{a^3}{2H}\int \frac{d^3k_1}{(2\pi)^3}\int \frac{d^3k_2}{(2\pi)^3}\int \frac{d^3k_3}{(2\pi)^3}(2\pi)^3\delta^{(3)} ({\bf k}_1+{\bf k}_2+{\bf k}_3)
\\
\label{zeta_phi2_normal}
&\quad
\times
\dot\zeta_{{\bf k}_3}
\left(\left(\dot{\varphi}_{k_1}\dot{\varphi}_{k_2}+m^2\varphi_{k_1}\varphi_{k_2}\right)a_{-{\bf k}_1}a_{-{\bf k}_2}
+\left(\dot{\varphi}_{k_1}^\ast\dot{\varphi}_{k_2}^\ast
+m^2\varphi_{k_1}^\ast\varphi_{k_2}^\ast\right)
a_{{\bf k}_1}^\dagger a_{{\bf k}_2}^\dagger
+2\left(\dot\varphi_{k_1}^\ast\dot\varphi_{k_2}
+m^2\varphi_{k_1}^\ast\varphi_{k_2}\right)
a_{{\bf k}_1}^\dagger a_{-{\bf k}_2}
\right)\,.
\end{align}
Here and in what follows,
we take the Hamiltonian in the normal-ordered form.
Using the mode function [Eq.~\eqref{mode_heavy_scalar}],
we obtain the relations
\begin{align}
\label{mode_heavy}
\dot{\varphi}_{k_1}\dot{\varphi}_{k_2}\simeq
-m^2\varphi_{k_1}\varphi_{k_2}
\simeq -\frac{m}{2a^3}e^{-2imt}\,,
\quad
\dot{\varphi}_{k_1}^\ast\dot{\varphi}_{k_2}^\ast
\simeq
-m^2\varphi_{k_1}^\ast\varphi_{k_2}^\ast
\simeq -\frac{m}{2a^3}e^{2imt}\,,
\quad
\dot{\varphi}_{k_1}^\ast\dot{\varphi}_{k_2}\simeq
m^2\varphi_{k_1}^\ast\varphi_{k_2}
\simeq \frac{m}{2a^3}\,,
\end{align}
which reduce Eq.~\eqref{zeta_phi2_normal} to the form
\begin{align}
\label{leading_zeta_phi}
\frac{a^3 }{2H}\int d^3x  \,\dot{\zeta}\Big(\dot{\phi}^2 + m^2\phi^2\Big)&\simeq\frac{m}{H}\int \frac{d^3k_1}{(2\pi)^3}\int \frac{d^3k_2}{(2\pi)^3}\int \frac{d^3k_3}{(2\pi)^3}(2\pi)^3\delta^{(3)} ({\bf k}_1+{\bf k}_2+{\bf k}_3)\,
\dot\zeta_{{\bf k}_3}
a_{{\bf k}_1}^\dagger a_{-{\bf k}_2}\,.
\end{align}
Note that the interaction in Eq.~\eqref{leading_zeta_phi}
represents the absorption/emission of $\zeta$
by the axion.
Similarly,
we rewrite the second term in Eq.~\eqref{eq4-7} as
\begin{align}
\nonumber
&-\frac{3a^3 }{2}\int d^3x  \,\zeta\Big(\dot{\phi}^2 -m^2\phi^2\Big)\\
\label{zeta_phi_pair}
&=\frac{3m}{2}\int \frac{d^3k_1}{(2\pi)^3}\int \frac{d^3k_2}{(2\pi)^3}\int \frac{d^3k_3}{(2\pi)^3}(2\pi)^3\delta^{(3)} ({\bf k}_1+{\bf k}_2+{\bf k}_3)\,
\zeta_{{\bf k}_3}
\left(e^{-2imt}a_{-{\bf k}_1}a_{-{\bf k}_2}
+e^{2imt}a_{{\bf k}_1}^\dagger a_{{\bf k}_2}^\dagger
\right)\,,
\end{align}
which represents
the pair creation/annihilation processes
of axions.
The magnitudes of the interactions in Eqs.~\eqref{leading_zeta_phi}
and~\eqref{zeta_phi_pair}
seem to be of the same order.
However,
the latter contains 
rapidly oscillating components~$e^{\pm 2imt}$,
so that its effect is suppressed by a factor of $H/m$
compared with the former.
An intuitive interpretation
of this suppression
is that the pair creation/annihilation processes
of heavy particles 
are rare compared with the absorption/emission
of light particles by heavy particles.
We then conclude that,
in the regime $m\gg k/a,H$,
the leading contribution from $\zeta\phi^2$-type interactions is given by
\begin{align}
\label{relevant_cubic}
H_{{\rm int},\zeta\phi^2}
&\simeq\frac{m}{H}\int \frac{d^3k_1}{(2\pi)^3}\int \frac{d^3k_2}{(2\pi)^3}\int \frac{d^3k_3}{(2\pi)^3}(2\pi)^3\delta^{(3)} ({\bf k}_1+{\bf k}_2+{\bf k}_3)\,
\dot\zeta_{{\bf k}_3}
a_{{\bf k}_1}^\dagger a_{-{\bf k}_2}\,.
\end{align}
Finally,
let us calculate
the leading contribution
from $\phi^4$-type interactions.
It is now straightforward
to extend the above discussions
to this type of interaction [Eq.~\eqref{eq4-8}].
In the regime $m\gg k/a,H$,
we have
\begin{align}
\label{relevant_phi4}
H_{{\rm int},\phi^4} \simeq
\frac{m^2}{4M_{\rm Pl}^2a^3H^2\tilde{\epsilon}}
\int \frac{d^3k_1}{(2\pi)^3}\int \frac{d^3k_2}{(2\pi)^3}
\int \frac{d^3k_3}{(2\pi)^3}\int \frac{d^3k_4}{(2\pi)^3}
(2\pi)^3\delta^{(3)} ({\bf k}_1+{\bf k}_2+{\bf k}_3+{\bf k}_4)
a_{{\bf k}_1}^\dagger a_{{\bf k}_2}^\dagger a_{-{\bf k}_3}a_{-{\bf k}_4}\,,
\end{align}
where we have used the relations in Eq.~\eqref{mode_heavy}.
Note that the relevant interaction [Eq.~\eqref{relevant_phi4}]
represents processes preserving the axion number.
In the next subsection,
we evaluate the effective scalar quartic interaction
induced by these interactions [Eqs.~\eqref{relevant_cubic} and~\eqref{relevant_phi4}]
relevant in the regime $m\gg k/a,H$.

\subsection{\label{sec3-2}Effective scalar quartic interaction}
\begin{figure}[htbp]
\begin{center}
\includegraphics[scale=0.7]{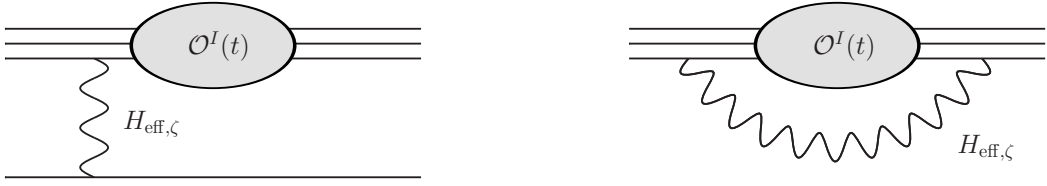}
\end{center}
\caption{Schematics of the terms appearing in Eq.~\eqref{eq3-normordered}.
The blob represents some operator $\mathcal{O}^I(t)$, which consists of a certain set of
$a_{\bf k}$'s and $a_{\bf k}^\dagger$'s. For instance, $\mathcal{O}^I(t)$ contains three
$a_{\bf k}$'s and three $a_{\bf k}^\dagger$'s in these figures.
The effective interaction $H_{{\rm eff},\zeta}$ can be identified as four legs connected by a wavy line,
which represents the propagation of $\zeta$.
A leg should be absorbed by the blob if there exists a contraction such as $\left[a_{\bf k},\mathcal{O}^I(t)\right]$
or $\left[a^{\dagger}_{\bf k},\mathcal{O}^I(t)\right]$.
The left figure shows the first two terms in the brackets, which correspond to tree-level diagrams.
The right figure shows the last two terms, which correspond to loop diagrams.}
\label{fig2}
\end{figure}

Let us calculate the expectation value of
some operator $\mathcal{O}(t)$
using the in-in formalism~\cite{Weinberg:2005vy}:
\begin{align}
\label{in-in_general}
\langle{\rm in}|\mathcal{O}(t)|{\rm in}\rangle
=\langle{\rm in}|
\left[\bar{T}\exp \Big(i\int_{t_0}^tdt^\prime H_{\rm int}(t^\prime)\Big)\right]
\mathcal{O}^I(t)
\left[T\exp \Big(-i\int_{t_0}^tdt^\prime H_{\rm int}(t^\prime)\Big)\right]
|{\rm in}\rangle\,,
\end{align}
where 
$T$~($\bar{T}$) denote (anti)time ordering,
$\mathcal{O}^I$ is constructed from the interaction picture fields,
and the interaction Hamiltonian $H_{\rm int}$
is given by $H_{\rm int}=H_{{\rm int},\zeta\phi^2}+H_{{\rm int},\phi^4}$
with Eqs.~\eqref{relevant_cubic} and \eqref{relevant_phi4}.
At the leading order
in the gravitational coupling,
Eq.~\eqref{in-in_general} is expanded as
\begin{align}
\nonumber
\langle{\rm in}|\mathcal{O}(t)|{\rm in}\rangle
=  
\langle{\rm in}|\mathcal{O}^I(t)|{\rm in}\rangle
&+ i\int^t_{t_0}dt_1\langle{\rm in}|[H_{{\rm int},\phi^4}(t_1),\mathcal{O}^I(t)]|{\rm in}\rangle\\
&
+ i^2\int^t_{t_0}dt_2 \int^{t_2}_{t_0}dt_1
\langle{\rm in}|\left[H_{{\rm int},\zeta\phi^2}(t_1),\left[H_{{\rm int},\zeta\phi^2}(t_2),\mathcal{O}^I(t)\right]\right]|{\rm in}\rangle \,,
\label{in-in_general_leading}
\end{align}
where the last term contains $\zeta$-exchanging processes.
Note that,
for the tree-level calculation,
we further drop contributions corresponding to
loop diagrams.
We expect that,
when the operator $\mathcal{O}(t)$
does not contain the creation/annihilation operators
of $\zeta$, and the in-state $|{\rm in}\rangle$
satisfies $a_{z,{\bf k}}|{\rm in}\rangle=0$,
the tree-level calculation of Eq.~\eqref{in-in_general_leading}
can be reduced~to
\begin{align}
\langle{\rm in}|\mathcal{O}(t)|{\rm in}\rangle
&=\langle{\rm in}|\mathcal{O}^I(t)|{\rm in}\rangle
+ i\int^t_{t_0}dt_1\langle{\rm in}|[H_{{\rm eff}}(t_1),\mathcal{O}^I(t)]|{\rm in}\rangle\,, \label{in-in_tree}
\end{align}
where $H_{\rm eff}$ is some effective interaction Hamiltonian
constructed only from
the creation/annihilation operators of $\phi$.
In the following,
we construct the effective Hamiltonian $H_{\rm eff}$
in the form
\begin{align}
H_{\rm eff}=
H_{{\rm int},\phi^4}+H_{{\rm eff},\zeta}\,,
\end{align}
where the second term $H_{{\rm eff},\zeta}$ satisfies
the following relation at the tree level:
\begin{align}
\int^t_{t_0}dt_1
\langle{\rm in}|
[H_{{\rm eff},\zeta}(t_1),\mathcal{O}^I(t)]
|{\rm in}\rangle
&=i\int^t_{t_0}dt_2 \int^{t_2}_{t_0}dt_1
\langle{\rm in}|\left[H_{{\rm int},\zeta\phi^2}(t_1),\left[H_{{\rm int},\zeta\phi^2}(t_2),\mathcal{O}^I(t)\right]\right]|{\rm in}\rangle \,.
\label{eff_zeta}
\end{align}
To determine $H_{{\rm eff},\zeta}$,
let us calculate the right-hand side of Eq.~\eqref{eff_zeta}.
It is convenient to rewrite the commutation relation
$\left[H_{{\rm int},\zeta\phi^2}(t_2),\mathcal{O}^I(t)\right]$
as
\begin{align}
\nonumber
&\left[H_{{\rm int},\zeta\phi^2}(t_2),\mathcal{O}^I(t)\right]
\\
&=
\frac{m}{H(t_2)}\int \frac{d^3k_1}{(2\pi)^3}\int \frac{d^3k_2}{(2\pi)^3}\int \frac{d^3k_3}{(2\pi)^3}(2\pi)^3\delta^{(3)} ({\bf k}_1+{\bf k}_2+{\bf k}_3)\,
\dot\zeta_{{\bf k}_3}(t_2)
\bigg(
\left[a_{{\bf k}_1}^\dagger ,\mathcal{O}^I(t)\right]
a_{-{\bf k}_2}
+a_{{\bf k}_1}^\dagger
\left[a_{-{\bf k}_2},\mathcal{O}^I(t)\right]
\bigg)\,.
\end{align}
Similarly,
we rewrite the commutator
$\left[H_{{\rm int},\zeta\phi^2}(t_1),\left[H_{{\rm int},\zeta\phi^2}(t_2),\mathcal{O}^I(t)\right]\right]$
as
\begin{align}
\nonumber
&\left[H_{{\rm int},\zeta\phi^2}(t_1),\left[H_{{\rm int},\zeta\phi^2}(t_2),\mathcal{O}^I(t)\right]\right]\\
\nonumber
&\qquad
=
\frac{m}{H(t_1)}
\int \frac{d^3k_1}{(2\pi)^3}\int \frac{d^3k_2}{(2\pi)^3}\int \frac{d^3k_3}{(2\pi)^3}(2\pi)^3\delta^{(3)} ({\bf k}_1+{\bf k}_2+{\bf k}_3)\\
&\qquad\qquad\quad
\times
\left(
\left[\dot\zeta_{{\bf k}_3}(t_1),\left[H_{{\rm int},\zeta\phi^2}(t_2),\mathcal{O}^I(t)\right]\right]
a_{{\bf k}_1}^\dagger a_{-{\bf k}_2}
+
\dot\zeta_{{\bf k}_3}(t_1)
\left[
a_{{\bf k}_1}^\dagger a_{-{\bf k}_2}
,\left[H_{{\rm int},\zeta\phi^2}(t_2),\mathcal{O}^I(t)\right]\right]
\right)\,.
\label{H_zetaphi2_O}
\end{align}
Then,
the first term in the parentheses
can be written as
\begin{align}
\nonumber
&\left[\dot\zeta_{{\bf k}_3}(t_1),\left[H_{{\rm int},\zeta\phi^2}(t_2),\mathcal{O}^I(t)\right]\right]
a_{{\bf k}_1}^\dagger a_{-{\bf k}_2}
\\
\nonumber
&=
\frac{m}{H(t_2)}\int \frac{d^3k_4}{(2\pi)^3}\int \frac{d^3k_5}{(2\pi)^3}\int \frac{d^3k_6}{(2\pi)^3}(2\pi)^3\delta^{(3)}({\bf k}_4+{\bf k}_5+{\bf k}_6)\\
&\qquad\qquad
\times\left[\dot\zeta_{{\bf k}_3}(t_1),\dot\zeta_{{\bf k}_6}(t_2)\right]
\bigg(
\left[a_{{\bf k}_4}^\dagger ,\mathcal{O}^I(t)\right]
a_{-{\bf k}_5}
+a_{{\bf k}_4}^\dagger
\left[a_{-{\bf k}_5},\mathcal{O}^I(t)\right]
\bigg)
a_{{\bf k}_1}^\dagger a_{-{\bf k}_2}\,.
\end{align}
Using $\left[\dot\zeta_{{\bf k}_3}(t_1),\dot\zeta_{{\bf k}_6}(t_2)\right]=
(2\pi)^3\delta^{(3)}({\bf k}_3+{\bf k}_6)\Big(\dot{\mathcal{Z}}_{k_3}^\ast(t_2)\dot{\mathcal{Z}}_{k_3}(t_1)
-\dot{\mathcal{Z}}_{k_3}(t_2)\dot{\mathcal{Z}}_{k_3}^\ast(t_1)\Big)$
and rewriting $a_{\bf k}$'s and $a_{\bf k}^\dagger$'s
in the normal-ordered form,
we obtain
\begin{align}
\nonumber
&\left[\dot\zeta_{{\bf k}_3}(t_1),\left[H_{{\rm int},\zeta\phi^2}(t_2),\mathcal{O}^I(t)\right]\right]
a_{{\bf k}_1}^\dagger a_{-{\bf k}_2}
\\
\nonumber
&=
\frac{m}{H(t_2)}\int \frac{d^3k_4}{(2\pi)^3}\int \frac{d^3k_5}{(2\pi)^3}(2\pi)^3\delta^{(3)}({\bf k}_4+{\bf k}_5-{\bf k}_3)
\\
&
\nonumber
\qquad
\times
\Big(\dot{\mathcal{Z}}_{k_3}^\ast(t_2)\dot{\mathcal{Z}}_{k_3}(t_1)
-\dot{\mathcal{Z}}_{k_3}(t_2)\dot{\mathcal{Z}}_{k_3}^\ast(t_1)\Big)
\Bigg[
a_{{\bf k}_1}^\dagger\left[a_{{\bf k}_4}^\dagger ,\mathcal{O}^I(t)\right]
a_{-{\bf k}_5}a_{-{\bf k}_2}
+a_{{\bf k}_1}^\dagger a_{{\bf k}_4}^\dagger
\left[a_{-{\bf k}_5},\mathcal{O}^I(t)\right]a_{-{\bf k}_2}
\\
&\qquad\qquad\qquad
\qquad\qquad\qquad\qquad
\qquad\qquad\quad
-
\left[a_{{\bf k}_1}^\dagger,\left[a_{{\bf k}_4}^\dagger ,\mathcal{O}^I(t)\right]
a_{-{\bf k}_5}\right]a_{-{\bf k}_2}
-\left[a_{{\bf k}_1}^\dagger,a_{{\bf k}_4}^\dagger
\left[a_{-{\bf k}_5},\mathcal{O}^I(t)\right]\right]a_{-{\bf k}_2}
\Bigg]
\,.\label{eq3-normordered}
\end{align}
As depicted in Fig.~\ref{fig2},
the first two terms in the brackets
correspond to tree-level diagrams with the $\zeta$ exchange,
and the last two terms
correspond to loop diagrams.
We therefore drop the last two terms
in the tree-level calculation.
Similarly,
the second term in the parentheses in Eq.~\eqref{H_zetaphi2_O}
contains double contractions of $a_{\bf k}$
and $a_{\bf k}^\dagger$,
so it corresponds to loop-level diagrams,
and it is irrelevant in the tree level calculation.
Based on the above discussions,
we conclude that
the right-hand side of Eq.~\eqref{eff_zeta}
can be calculated at the tree level as
\begin{align}
\nonumber
&i\int^t_{t_0}dt_2 \int^{t_2}_{t_0}dt_1
\langle{\rm in}|\left[H_{{\rm int},\zeta\phi^2}(t_1),\left[H_{{\rm int},\zeta\phi^2}(t_2),\mathcal{O}^I(t)\right]\right]|{\rm in}\rangle\\
\nonumber
&=i\int^t_{t_0}dt_2 \int^{t_2}_{t_0}dt_1\int \frac{d^3k_1}{(2\pi)^3}\int \frac{d^3k_2}{(2\pi)^3}
\int \frac{d^3k_3}{(2\pi)^3}\int \frac{d^3k_4}{(2\pi)^3}
(2\pi)^3\delta^{(3)} ({\bf k}_1+{\bf k}_2-{\bf k}_3-{\bf k}_4)
\\
\nonumber
&\qquad
\times
\frac{m^2}{H(t_1)H(t_2)}
\Big(\dot{\mathcal{Z}}_{|{\bf k}_2-{\bf k}_3|}^\ast(t_2)\dot{\mathcal{Z}}_{|{\bf k}_2-{\bf k}_3|}(t_1)
-\dot{\mathcal{Z}}_{|{\bf k}_2-{\bf k}_3|}(t_2)\dot{\mathcal{Z}}_{|{\bf k}_2-{\bf k}_3|}^\ast(t_1)\Big)\\
&\qquad
\times\langle{\rm in}|
\left(
a_{{\bf k}_1}^\dagger\left[a_{{\bf k}_2}^\dagger ,\mathcal{O}^I(t)\right]
a_{{\bf k}_3}a_{{\bf k}_4}
+a_{{\bf k}_1}^\dagger a_{{\bf k}_2}^\dagger
\left[a_{{\bf k}_3},\mathcal{O}^I(t)\right]a_{{\bf k}_4}
\right)
|{\rm in}\rangle\,, \label{H_eff_tree_RHS}
\end{align}
where we relabeled the momentum variables such that ${\bf k}_4\to{\bf k}_2$, ${\bf k}_5\to-{\bf k}_3$, ${\bf k}_2\to-{\bf k}_4$.
Note that we can replace the index $|{\bf k}_2-{\bf k}_3|$ in the mode function $\mathcal{\dot{Z}}$ in Eq.~\eqref{H_eff_tree_RHS} with
$|{\bf k}_1-{\bf k}_3|$ by using the delta function $\delta^{(3)} ({\bf k}_1+{\bf k}_2-{\bf k}_3-{\bf k}_4)$
and symmetries for two of the $a_{\bf k}$'s and $a_{\bf k}^\dagger$'s:
$a_{{\bf k}_3}a_{{\bf k}_4}=a_{{\bf k}_4}a_{{\bf k}_3}$ and $a^{\dagger}_{{\bf k}_1}a^{\dagger}_{{\bf k}_2}=a^{\dagger}_{{\bf k}_2}a^{\dagger}_{{\bf k}_1}$.
It is now easy to determine $H_{{\rm eff},\zeta}$ as
\begin{align}
\nonumber
H_{{\rm eff},\zeta}(t)&=
\frac{i}{2}\int^{t}_{t_0}dt^\prime\int \frac{d^3k_1}{(2\pi)^3}\int \frac{d^3k_2}{(2\pi)^3}
\int \frac{d^3k_3}{(2\pi)^3}\int \frac{d^3k_4}{(2\pi)^3}
(2\pi)^3\delta^{(3)} ({\bf k}_1+{\bf k}_2-{\bf k}_3-{\bf k}_4)\\
&\qquad
\times
\frac{m^2}{H(t)H(t^\prime)}
\Big(\dot{\mathcal{Z}}_{|{\bf k}_1-{\bf k}_3|}^\ast(t)\dot{\mathcal{Z}}_{|{\bf k}_1-{\bf k}_3|}(t^\prime)
-\dot{\mathcal{Z}}_{|{\bf k}_1-{\bf k}_3|}(t)\dot{\mathcal{Z}}_{|{\bf k}_1-{\bf k}_3|}^\ast(t^\prime)\Big)
a_{{\bf k}_1}^\dagger a_{{\bf k}_2}^\dagger a_{{\bf k}_3}a_{{\bf k}_4}\,,
\end{align}
which reproduces the relation~\eqref{eff_zeta}, if we ignore double contractions such as $\left[\left[a^{\dagger}_{{\bf k}_1},\mathcal{O}^I(t)\right],a^{\dagger}_{{\bf k}_2}\right]a_{{\bf k}_3}a_{{\bf k}_4}$, etc.
Then,
the total effective interaction Hamiltonian $H_{\rm eff}$
is given~by
\begin{align}
H_{\rm eff}(t)
&=\int \frac{d^3k_1}{(2\pi)^3}\int \frac{d^3k_2}{(2\pi)^3}
\int \frac{d^3k_3}{(2\pi)^3}\int \frac{d^3k_4}{(2\pi)^3}
(2\pi)^3\delta^{(3)} ({\bf k}_1+{\bf k}_2-{\bf k}_3-{\bf k}_4)
F(t;|{\bf k}_1-{\bf k}_3|)
a_{{\bf k}_1}^\dagger a_{{\bf k}_2}^\dagger a_{{\bf k}_3}a_{{\bf k}_4}\,, \label{total_Heff}
\end{align}
where the function $F(t;k)$ is defined as
\begin{align}
F(t;k)&=\frac{m^2}{4M_{\rm Pl}^2a^3H^2\tilde{\epsilon}}
+\frac{i}{2}\int_{t_0}^tdt^\prime
\frac{m^2}{H(t)H(t^\prime)}
\Big(\dot{\mathcal{Z}}_{k}^\ast(t)\dot{\mathcal{Z}}_{k}(t^\prime)
-\dot{\mathcal{Z}}_{k}(t)\dot{\mathcal{Z}}_{k}^\ast(t^\prime)\Big).
\end{align}
In the next subsection
we calculate the function $F(t;k)$ explicitly
for the radiation-dominated universe.

\subsection{\label{sec3-3}Application to radiation-dominated universe}
Finally,
we apply the results in the previous subsections
to the radiation-dominated universe.
For the radiation-dominated universe,
$a(t)\propto t^{1/2}$,
and the Hubble parameter $H$
and the parameter $\tilde{\epsilon}$
are given by
\begin{align}
H(t)=\frac{1}{2t}\,,
\quad
\tilde{\epsilon}=2c_s^{-2}\,.
\end{align}
Furthermore, the mode function $\mathcal{Z}_{k}$ is given by
(see Appendix~\ref{secB} for details)
\begin{align}
\mathcal{Z}_{k}(t)
=\frac{c_s}{2\sqrt{2}aM_{\rm Pl}}\frac{1}{\sqrt{c_sk}}e^{-ic_sk\tau}\,,
\end{align}
where the conformal time $\tau$
satisfies $\tau=1/(aH)\propto t^{1/2}$.
Then, the function $F(t;k)$ takes the form
\begin{align}
F(t;k)&=
\frac{c_s^{2}m^2}{8M_{\rm Pl}^2a^3H^2}
-\frac{c_s^{2}m^2}{8M_{\rm Pl}^2}\frac{1}{c_sk}
\int_{0}^tdt^\prime
\frac{1}{H(t)H(t^\prime)}
\frac{d^2}{dtdt^\prime}\left[
\frac{1}{a(t)a(t^\prime)}\sin c_sk(\tau-\tau^\prime)
\right]\,,
\label{f_RD}
\end{align}
where we set $t_0=0$, assuming that
the in-state is defined at a sufficiently early time.
It is not difficult to perform the integral
in Eq.~\eqref{f_RD},
and we obtain
\begin{align}
F(t;k)=-\frac{m^2}{4M_{\rm Pl}^2a^3}\frac{1}{(k/a)^2}\,
f\Big(\frac{c_sk/a}{H}\Big)
=-\frac{2\pi Gm^2}{a^3}\frac{1}{(k/a)^2}\,
f\Big(\frac{c_sk/a}{H}\Big)
\quad
{\rm with}
\quad
f(x)=1-\cos x-x\sin x\,.
\end{align}
Here, it should be noted that
the function $F(t;k)$
depends on the ratio of the momentum transfer $k/a$
and the sound horizon $H/c_s$,
and the function $f(x)$ behaves as
\begin{align}
f(x)=\left\{\begin{array}{ccc}
1+(\text{highly oscillating terms})&\quad{\rm for}\quad&x\gg1\,,\\[2mm]
\displaystyle-\frac{1}{2}x^2&\quad{\rm for}\quad&x\ll1\,.
\end{array}\right. \label{fxlimit}
\end{align}
We then notice that
the effective quartic interaction
reproduces the Newtonian approximation
when the momentum transfer
is subhorizon scale, $c_sk/(aH)\gg1$:
\begin{align}
H_{\rm eff}\simeq
\int \frac{d^3k_1}{(2\pi)^3}\int \frac{d^3k_2}{(2\pi)^3}
\int \frac{d^3k_3}{(2\pi)^3}\int \frac{d^3k_4}{(2\pi)^3}
(2\pi)^3\delta^{(3)} ({\bf k}_1+{\bf k}_2-{\bf k}_3-{\bf k}_4)
\left[-\frac{2\pi Gm^2}{a^3}\frac{1}{(|{\bf k}_1-{\bf k}_3|/a)^2}\right]
a_{{\bf k}_1}^\dagger a_{{\bf k}_2}^\dagger a_{{\bf k}_3}a_{{\bf k}_4}\,,
\end{align}
where we have dropped highly oscillating terms.
On the other hand,
when the momentum transfer
is superhorizon scale, $c_sk/(aH)\ll1$,
the effective interaction
is saturated by the sound horizon scale $H/c_s$ as
\begin{align}
H_{\rm eff}\simeq
\int \frac{d^3k_1}{(2\pi)^3}\int \frac{d^3k_2}{(2\pi)^3}
\int \frac{d^3k_3}{(2\pi)^3}\int \frac{d^3k_4}{(2\pi)^3}
(2\pi)^3\delta^{(3)} ({\bf k}_1+{\bf k}_2-{\bf k}_3-{\bf k}_4)
\left[\frac{\pi Gm^2}{a^3}\frac{1}{(H/c_s)^2}\right]
a_{{\bf k}_1}^\dagger a_{{\bf k}_2}^\dagger a_{{\bf k}_3}a_{{\bf k}_4}\,,
\label{eq3-3-Heffout}
\end{align}
which is suppressed by a factor $(c_sk/a)^2/H^2$
compared with the Newtonian approximation.
As is expected,
our results based on the general relativistic framework
reproduced the Newtonian approximation
at the subhorizon scale
and we also confirmed the 
saturation
of gravitational interactions
at the superhorizon scale.
In the next section
we apply our results to the system of coherently oscillating axions
and discuss its implications for axion cosmology.

\section{\label{sec4} Self-interaction of axions in coherent states}
Now that we have determined the interaction Hamiltonian, we can consider the implications for axion models and cosmology.
Before going into the discussion on the gravitational interactions of dark matter axions, we note that the behavior of axions
is closely related to the history of the early universe [see Refs.~\cite{Sikivie:2006ni,Kawasaki:2013ae} for reviews on axion cosmology].
The axion arises as a Goldstone boson when Peccei-Quinn (PQ) symmetry is spontaneously broken~\cite{Weinberg:1977ma},
and it remains massless until the time of the QCD phase transition (let us denote this time as $t=t_q$), at which the axion acquires a mass.
After acquiring the mass, when $m\gtrsim H$ is satisfied, the axion field begins to oscillate around the minimum of its potential, which is called the misalignment production
mechanism~\cite{Preskill:1982cy}.
However, the composition of dark matter axions varies depending on whether PQ symmetry is broken after inflation or not.
If PQ symmetry is broken after inflation, topological defects such as strings and domain walls are formed, and their annihilation also produces axions in addition to
those produced by the misalignment mechanism~\cite{Davis:1986xc}.
On the other hand, if PQ symmetry is broken before the end of inflation, dark matter axions are just produced by the misalignment mechanism.

As in the previous work~\cite{Saikawa:2012uk}, we describe the axions produced by the misalignment mechanism as coherent states~\cite{Glauber:1963tx}
of the axion field. The coherent state description is applicable to the modes produced outside the horizon, since it has the same trajectory as the classical field.
Strictly speaking, the coherent state $|\{\alpha\}\rangle$ is defined as a state which factorizes $n$th-order correlation functions in terms of a single function which is an eigenvalue of the field operator~\cite{Glauber:1963fi,Glauber:1963tx}:
\begin{equation}
\langle\{\alpha\}|\phi^{(-)}(x_1)\cdots\phi^{(-)}(x_n)\phi^{(+)}(x_{n+1})\cdots\phi^{(+)}(x_{2n})|\{\alpha\}\rangle = \prod^n_{i=1}\Phi^*(x_i)\prod_{j=n+1}^{2n}\Phi(x_j), \label{eq4-1}
\end{equation}
where
\begin{align}
&\phi^{(+)}(x) = \int \frac{d^3 k}{(2\pi)^3}e^{i{\bf k\cdot x}}\varphi_k(t)a_{\bf k}, \label{eq4-2} \\
&\phi^{(-)}(x) = \int \frac{d^3 k}{(2\pi)^3}e^{-i{\bf k\cdot x}}\varphi^*_k(t)a^{\dagger}_{\bf k}, \label{eq4-3} \\
&\phi^{(+)}(x)|\{\alpha\}\rangle = \Phi(x)|\{\alpha\}\rangle, \label{eq4-4}
\end{align}
and $a_{\bf k}$ is the annihilation operator of the axion field satisfying the commutation relations in Eq.~\eqref{eq2-commak}.
Equation~\eqref{eq4-1} implies that measurements at $2n$ different spacetime points are statistically independent,
and results are given by the product of the field amplitude $\Phi(x)$.
Coherent oscillation of free axions produced by the misalignment mechanism can be described in terms of the evolution of the function $\Phi(x)$~\cite{Saikawa:2012uk}.

On the other hand, we expect that axions produced by other mechanisms such as the thermal production~\cite{Turner:1986tb} and
the decay of topological defects~\cite{Davis:1986xc}
are not described by coherent states, since microscopic procedures lead to nontrivial correlation functions for the axion field, which violates the coherence condition [Eq.~\eqref{eq4-1}].
Assuming that these axions are described as number states, one can show that their interaction rate is negligible in the cosmological time scale~\cite{Saikawa:2012uk}.
Henceforth, we just concentrate on the self-interaction of axions in coherent states,
which are produced by the misalignment mechanism,
and simply ignore the contributions from other states.
As we will see below, the self-interactions between axions in coherent states might become relevant during the radiation-dominated era, but the interactions between those produced from other mechanisms (thermal bath and topological defects) would be irrelevant because of the reason described above. In particular, if PQ symmetry is broken after inflation, we expect that gravitational self-interactions affect only the behavior of a fraction of dark matter, since in this case the population of dark matter axions is dominated by those produced by topological defects~\cite{Davis:1986xc}.

It should be noted that axions in coherent states involve the modes with nonzero wave number $k$.
This is because the value of the axion field varies randomly over the horizon scale in the case where PQ symmetry is broken after inflation.
Furthermore, even in the case where PQ symmetry is broken before the end of inflation, the nonzero modes exist due to the quantum fluctuations of the
massless axion field during inflation. Among these nonzero modes, however, we ignore the modes which enter the horizon before the time of the QCD phase transition
$t=t_q$, since they begin to oscillate before $t_q$ and their amplitudes become smaller than that of the modes outside the horizon before $t_q$.
Therefore, we describe the axions in coherent states in terms of the modes outside the horizon at the time of the QCD phase transition
\begin{equation}
|\{\alpha\}\rangle = \prod_{k<H_qa_q}|\alpha_{\bf k}\rangle, \label{eq4-5}
\end{equation}
where $H_q$ and $a_q$ are the Hubble parameter and the scale factor at the time $t_q$, and $|\alpha_{\bf k}\rangle$ satisfies
\begin{equation}
a_{\bf k}|\alpha_{\bf k}\rangle = \sqrt{V}\alpha_{\bf k}|\alpha_{\bf k}\rangle, \label{eq4-6}
\end{equation}
with a c-number eigenvalue $\alpha_{\bf k}$. The factor $V$ represents the volume of the spatial box, 
for which we take a limit $V\to \infty$ after we complete the calculation~\cite{Saikawa:2012uk}.
The coherent state satisfying Eq.~\eqref{eq4-6} can be constructed in terms of the creation operators:
\begin{equation}
|\alpha_{\bf k}\rangle = e^{-\frac{1}{2}|\alpha_{\bf k}|^2}\sum_{n=0}^{\infty}\frac{\alpha_{\bf k}^n}{n!\sqrt{V^n}}(a_{\bf k}^{\dagger})^n|0\rangle,
\end{equation}
where the state $|0\rangle$ is defined by
\begin{equation}
a_{\bf k}|0\rangle = 0 \qquad {\rm and} \qquad a_{z{\bf k}}|0\rangle = 0.
\label{eq4-def_of_vac}
\end{equation}

Let us consider the gravitational self-interactions of the states given by Eq.~\eqref{eq4-5}.
We would like to estimate the interaction rate of axions due to the effective gravitational interactions derived in the previous section.
Here, the interaction rate $\Gamma$ is defined as the time scale in which the occupation number of axions changes its value.
Assuming that the interactions are absent at sufficiently early times, we can estimate the number of axions occupying
the state labeled by a (comoving) momentum ${\bf p}$ by using the number operator
\begin{equation}
\mathcal{N}_{\bf p} = \frac{1}{V}a^{\dagger}_{\bf p}a_{\bf p}, \label{eq4-def_of_Np}
\end{equation}
which diagonalizes the free Hamiltonian. In the above equation, the factor $1/V$ is multiplied in order to compensate the normalization of the creation
and annihilation operators. The time evolution of the expectation value of $\mathcal{N}_{\bf p}$ can be calculated by using the
formula~\eqref{eqB-1-4} with $\mathcal{O}^I = \mathcal{N}_{\bf p}$~\cite{Saikawa:2012uk}.
Here, we take the in-state as $|{\rm in}\rangle= |\{\alpha\}\rangle$, since we are interested in the self-interaction of axions in the coherent states.
Then it follows that
\begin{equation}
\langle\mathcal{N}_{\bf p}(t)\rangle =  \mathcal{N}_{\bf p}(t_0) + i\int^t_{t_0}dt_1\langle[H_{{\rm int}}(t_1),{\cal N}_{\bf p}]\rangle
+ i^2\int^t_{t_0}dt_2 \int^{t_2}_{t_0}dt_1\langle\left[H_{{\rm int}}(t_1),\left[H_{{\rm int}}(t_2),{\cal N}_{\bf p}\right]\right]\rangle + \dots, \label{eq4-10}
\end{equation}
where dots represent the terms of higher order in $H_{\rm int}$, and $\langle\dots\rangle$ denotes the expectation value
for the coherent states, $\langle\{\alpha\}|\dots|\{\alpha\}\rangle$.
Once we obtain the expectation value $\langle\mathcal{N}_{\bf p}(t)\rangle$, the interaction rate can be estimated as
\begin{equation}
\Gamma = \frac{1}{\langle\mathcal{N}_{\bf p}(t)\rangle}\frac{d\langle\mathcal{N}_{\bf p}(t)\rangle}{dt}. \label{eq4-9}
\end{equation}

In Ref.~\cite{Saikawa:2012uk}, it was shown that the term of the first order in the quartic interaction does not vanish as long as axions are in the condensed regime,
where their interaction rate is larger than the energy exchanged in the transition process.
Hence, it will be enough to calculate the first-order terms in the quartic interactions (or tree diagrams), since we are interested in the process occurring in the condensed regime.
In the previous section we learned that the following effective Hamiltonian can be used for tree level processes of gravitational interactions:
\begin{align}
H_{\rm eff}(t) = \int \frac{d^3 k_1}{(2\pi)^3}\int \frac{d^3 k_2}{(2\pi)^3}\int \frac{d^3 k_3}{(2\pi)^3}\int \frac{d^3 k_4}{(2\pi)^3}(2\pi)^3\delta^{(3)}({\bf k}_1+ {\bf k}_2 - {\bf k}_3 - {\bf k}_4)
F(t;|{\bf k}_1 - {\bf k}_3|)a^{\dagger}_{{\bf k}_1}a^{\dagger}_{{\bf k}_2}a_{{\bf k}_3}a_{{\bf k}_4}, \label{eq4-Heff} \\
F(t;k)= -\frac{2\pi Gm^2}{a^3(t)}\frac{a^2(t)}{k^2}f\left(  \frac{k}{k_H(t)}  \right) \qquad{\rm with} \qquad f(x) = 1- \cos x - x\sin x, \label{eq4-F} 
\end{align}
where $k_H(t) \equiv a(t)H(t) / c_s$.
Then, using Eq.~\eqref{in-in_tree} we immediately obtain
\begin{equation}
\langle\mathcal{N}_{\bf p}(t)\rangle \simeq \mathcal{N}_{\bf p}(t_0) + i\int^t_{t_0}dt_1\langle[H_{{\rm eff}}(t_1),{\cal N}_{\bf p}]\rangle \label{eq4-Nt}
\end{equation}
and 
\begin{align}
&i\int^t_{t_0}dt_1\langle[H_{{\rm eff}}(t_1),{\cal N}_{\bf p}]\rangle \nonumber\\
\quad&= i\int^t_{t_0}dt_1\frac{1}{V}\int \frac{d^3 k_1}{(2\pi)^3}\int \frac{d^3 k_2}{(2\pi)^3}\int \frac{d^3 k_3}{(2\pi)^3}\int \frac{d^3 k_4}{(2\pi)^3}\nonumber\\
&\qquad\times (2\pi)^3\delta^{(3)}({\bf k}_1+ {\bf k}_2 - {\bf k}_3 - {\bf k}_4)F(t_1;|{\bf k}_1 - {\bf k}_3|)
\langle[a^{\dagger}_{{\bf k}_1}a^{\dagger}_{{\bf k}_2}a_{{\bf k}_3}a_{{\bf k}_4}, a^{\dagger}_{\bf p}a_{\bf p}]\rangle \nonumber\\
\quad&= i\int^t_{t_0}dt_1\frac{2}{V}\int \frac{d^3 k_1}{(2\pi)^3}\int \frac{d^3 k_2}{(2\pi)^3}\int \frac{d^3 k_3}{(2\pi)^3}(2\pi)^3\delta^{(3)}({\bf k}_1+ {\bf k}_2 - {\bf k}_3 - {\bf p})F(t_1;|{\bf k}_1 - {\bf p}|)
\langle a^{\dagger}_{{\bf k}_1}a^{\dagger}_{{\bf k}_2}a_{{\bf k}_3}a_{\bf p} - a^{\dagger}_{\bf p}a^{\dagger}_{{\bf k}_3}a_{{\bf k}_2}a_{{\bf k}_1} \rangle \nonumber\\
\quad&= i\int^t_{t_0}dt_12V\int\frac{d^3 k_1}{(2\pi)^3}\bigg|_{k_1\lesssim H_qa_q}\int\frac{d^3 k_2}{(2\pi)^3}\bigg|_{k_2\lesssim H_qa_q}\int\frac{d^3 k_3}{(2\pi)^3}\bigg|_{k_3\lesssim H_qa_q} \nonumber\\
&\qquad\times (2\pi)^3\delta^{(3)}({\bf k}_1+ {\bf k}_2 - {\bf k}_3 - {\bf p}) F(t_1;|{\bf k}_1 - {\bf p}|) (\alpha^*_{{\bf k}_1}\alpha^*_{{\bf k}_2}\alpha_{{\bf k}_3}\alpha_{\bf p} - \alpha^*_{\bf p}\alpha^*_{{\bf k}_3}\alpha_{{\bf k}_2}\alpha_{{\bf k}_1}). \label{eq4-commutator}
\end{align}
The time derivative of this expression leads to
\begin{align}
\frac{d\langle\mathcal{N}_{\bf p}(t)\rangle}{dt} &= 2iV\int\frac{d^3 k_1}{(2\pi)^3}\bigg|_{k_1\lesssim H_qa_q}\int\frac{d^3 k_2}{(2\pi)^3}\bigg|_{k_2\lesssim H_qa_q}\int\frac{d^3 k_3}{(2\pi)^3}\bigg|_{k_3\lesssim H_qa_q} \nonumber\\
&\qquad\times (2\pi)^3\delta^{(3)}({\bf k}_1+ {\bf k}_2 - {\bf k}_3 - {\bf p})F(t;|{\bf k}_1 - {\bf p}|) 
(\alpha^*_{{\bf k}_1}\alpha^*_{{\bf k}_2}\alpha_{{\bf k}_3}\alpha_{\bf p} - \alpha^*_{\bf p}\alpha^*_{{\bf k}_3}\alpha_{{\bf k}_2}\alpha_{{\bf k}_1}). \label{eq4-dNdt}
\end{align}

The coefficient $\alpha_{\bf p}$ can be related to the field amplitude and the number for the mode ${\bf p}$ at the initial time.
From Eqs.~\eqref{eq4-2}-\eqref{eq4-4} and Eq.~\eqref{eq4-def_of_Np}, we obtain
\begin{align}
\langle\phi(t_0,{\bf x})\rangle &= \Phi(t_0,{\bf x}) + \Phi^*(t_0,{\bf x})
= \int\frac{d^3k}{(2\pi)^3}\left(e^{i{\bf k\cdot x}}\varphi_k(t_0)\alpha_{\bf k} + e^{-i{\bf k\cdot x}}\varphi_k^*(t_0)\alpha^*_{\bf k}\right), \label{eq4-17} \\
\langle \mathcal{N}_{\bf p} \rangle &= \mathcal{N}_{\bf p}(t_0) = |\alpha_{\bf p}|^2. \label{eq4-18}
\end{align}
In principle, the value of $\alpha_{\bf p}$ can be determined if we know the field configuration beyond the horizon scale at the time of the QCD phase transition $\langle\phi(t_0,{\bf x})\rangle$.
For instance, if PQ symmetry is broken after inflation, the field value of $\phi$ varies on a scale comparable to the QCD horizon $\sim H_q^{-1}$,
and hence the amplitude of $\alpha_{\bf p}$ changes almost randomly for the modes $p\lesssim H_qa_q$.
On the other hand, if PQ symmetry is broken before inflation, the field value $\langle\phi({\bf x})\rangle$ is homogenized due to the rapid expansion.
In this case, $\alpha_{\bf p=0}$ corresponds to the amplitude of a large homogeneous background, and
$\alpha_{\bf p\ne 0}$ corresponds to the amplitude of small perturbations, which originate from quantum fluctuations generated during inflation.
These fluctuations are regarded as isocurvature modes, whose magnitude is constrained below $\mathcal{O}(10^{-5})$ from recent observations~\cite{Beltran:2006sq}.
Therefore, we expect that the typical amplitude of $\alpha_{\bf p\ne 0}$ is suppressed by a factor of $\mathcal{O}(10^{-5})$ compared with that of $\alpha_{\bf p=0}$.
As we will discuss later, this fact leads to a modification on the estimation of the interaction rate.
We also note that the complex phase of $\alpha_{\bf p}$ does not vanish in general, which might affect the result of calculations.

Performing the integration over ${\bf k}_3$, we can write Eq.~\eqref{eq4-dNdt} as
\begin{equation}
\frac{d\langle\mathcal{N}_{\bf p}(t)\rangle}{dt} = 2iV\int\frac{d^3 k_1}{(2\pi)^3}\bigg|_{k_1\lesssim H_qa_q}\int\frac{d^3 k_2}{(2\pi)^3}\bigg|_{k_2\lesssim H_qa_q}
F(t;|{\bf k}_1 - {\bf p}|)(\alpha^*_{{\bf k}_1}\alpha^*_{{\bf k}_2}\alpha_{{\bf k}_1+{\bf k}_2-{\bf p}}\alpha_{\bf p} - \alpha^*_{\bf p}\alpha^*_{{\bf k}_1+{\bf k}_2-{\bf p}}\alpha_{{\bf k}_2}\alpha_{{\bf k}_1} ).
\label{eq4-dNdt2}
\end{equation}
From this expression, it immediately follows that the right-hand side of Eq.~\eqref{eq4-dNdt2} vanishes in the limit $|{\bf k}_1-{\bf p}|\to 0$,
since the coefficient containing $\alpha_{\bf p}$ becomes $\alpha_{\bf k_1}^*\alpha_{\bf k_2}^*\alpha_{{\bf k_1+k_2-p}}\alpha_{\bf p} \to |\alpha_{\bf k_2}|^2|\alpha_{\bf p}|^2$,
which cancels its complex conjugate. This fact just implies that there is no transition in the absence of the momentum transfer.
Furthermore, even though the value of $\alpha_{\bf p}$ changes almost randomly with ${\bf p}$, we intuitionally expect that $\alpha_{\bf p}$ would be (at least locally) analytic at a given value of ${\bf p}$, because it determines the amplitude of the field value $\langle\phi\rangle$.
In other words, we assume that for a sufficiently small interval $\Delta{\bf p}$ it can be expanded as
\begin{equation}
\alpha_{{\bf p}+\Delta{\bf p}} \simeq \alpha_{\bf p} + \frac{d\alpha_{\bf p}}{d{\bf p}}\cdot\Delta{\bf p} + \dots, \label{eq4-alphaexpansion}
\end{equation}
where the dots represent the terms of higher order in $\Delta{\bf p}$.
With this assumption, we deduce that
the transition rate is suppressed by the power of $|{\bf k}_1-{\bf p}|$, if the momentum transfer is sufficiently small.

\begin{figure}[htbp]
\begin{center}
\includegraphics[scale=1.0]{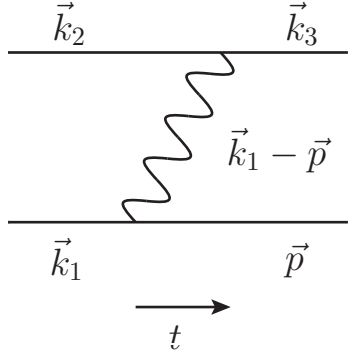}
\end{center}
\caption{Schematics of the quartic interaction contributing to Eq.~\eqref{eq4-dNdt}.}
\label{fig3}
\end{figure}

Now, let us look closely into the structure of Eq.~\eqref{eq4-dNdt}. What we aim to cope with in the present analysis based on general relativity is to make some distinction
between the processes involving modes inside and outside the (sound) horizon.
For this purpose, it would be convenient to decompose the integral over $k_1$, $k_2$, and $k_3$ in Eq.~\eqref{eq4-dNdt} into two parts:
\begin{equation}
\int\frac{d^3 k}{(2\pi)^3}\bigg|_{k\lesssim H_qa_q} \to \int\frac{d^3 k}{(2\pi)^3}\bigg|_{k< k_H(t)} + \int\frac{d^3 k}{(2\pi)^3}\bigg|_{k_H(t) < k \lesssim H_qa_q}. \label{eq4-dec}
\end{equation}
Also, for the wave number ${\bf p}$ appearing in Eq.~\eqref{eq4-dNdt}, we can consider two possibilities according to whether $p$ is greater than $k_H(t)$ or not.
Since the process contributing to the right-hand side of Eq.~\eqref{eq4-dNdt} can be written as a t-channel diagram, shown in Fig~\ref{fig3},
for quartic gravitational interactions of axions, we can consider the following three classes of interactions:
\begin{enumerate}
\item All external lines attached to the quartic coupling are subhorizon sized ($k_1,k_2,k_3,p> k_H$).
\item One or two external lines are superhorizon sized, while the other external lines and the momentum
transfer $|{\bf k}_1-{\bf p}|$ are subhorizon sized ($k_1,k_2> k_H$ and $k_3, p < k_H$, etc.).
\item All external lines are superhorizon sized ($k_1,k_2,k_3,p < k_H$).
\end{enumerate}
Among these three cases, we are particularly interested in case 2, since it represents a transition from (into) modes with higher momentum (subhorizon-sized waves)
into (from) almost homogeneous modes (superhorizon-sized waves), which is relevant to the transition into (from) BEC.

With the decomposition in Eq.~\eqref{eq4-dec}, Eq.~\eqref{eq4-dNdt} can be rewritten as
\begin{align}
\frac{d\langle\mathcal{N}_{\bf p}(t)\rangle}{dt} \bigg|_{p > k_H(t)} &= \frac{d\langle\mathcal{N}_{\bf p}(t)\rangle}{dt} \bigg|_1 + \frac{d\langle\mathcal{N}_{\bf p}(t)\rangle}{dt} \bigg|_{2,p > k_H(t)}, \label{eq4-dNdt12} \\
\frac{d\langle\mathcal{N}_{\bf p}(t)\rangle}{dt} \bigg|_{p < k_H(t)} &= \frac{d\langle\mathcal{N}_{\bf p}(t)\rangle}{dt} \bigg|_{2,p < k_H(t)} + \frac{d\langle\mathcal{N}_{\bf p}(t)\rangle}{dt} \bigg|_3, \label{eq4-dNdt23}
\end{align}
where the subscripts ``1," ``2," and ``3" are assigned to distinguish the terms contributing to the three cases enumerated above. For instance,
\begin{align}
\frac{d\langle\mathcal{N}_{\bf p}(t)\rangle}{dt}\bigg|_1 &= 2iV\int\frac{d^3 k_1}{(2\pi)^3}\bigg|_{k_H(t) < k_1\lesssim H_qa_q}\int\frac{d^3 k_2}{(2\pi)^3}\bigg|_{k_H(t) < k_2\lesssim H_qa_q}\int\frac{d^3 k_3}{(2\pi)^3}\bigg|_{k_H(t) < k_3\lesssim H_qa_q} \nonumber\\
&\qquad\times (2\pi)^3\delta^{(3)}({\bf k}_1+ {\bf k}_2 - {\bf k}_3 - {\bf p})F(t;|{\bf k}_1 - {\bf p}|) 
(\alpha^*_{{\bf k}_1}\alpha^*_{{\bf k}_2}\alpha_{{\bf k}_3}\alpha_{\bf p} - \alpha^*_{\bf p}\alpha^*_{{\bf k}_3}\alpha_{{\bf k}_2}\alpha_{{\bf k}_1}). \label{eq4-dNdt_1}
\end{align}
Similarly, there exist eight terms emerging from the decomposition [Eq.~\eqref{eq4-dec}], and 16 terms in total when we consider two possibilities for $p$.
We summarize the classification of these terms in Table~\ref{tab1}.
Note that there are some exceptions indicated as ``Others" in Table~\ref{tab1}: First,
the combinations [$k_2, k_3 < k_H(t) < k_1, p$]
and [$k_1, p < k_H(t) < k_2, k_3$]
are forbidden due to the conservation of three momenta, unless the momentum transfer $|{\bf k}_1-{\bf p}|$
becomes extremely small. As was shown below Eq.~\eqref{eq4-dNdt2}, such contributions are suppressed by the power of $|{\bf k}_1-{\bf p}|$.
Second, the combinations 
[$ k_1, k_2, k_3 < k_H(t) < p $],
[$ k_2, k_3, p < k_H(t) < k_1 $],
[$ k_1, k_3, p < k_H(t) < k_2 $],
and [$ k_1, k_2, p < k_H(t) < k_3$]
are also forbidden due to the conservation of three momenta 
except for the contributions involving the modes whose wavelength is comparable to the horizon.
These processes are not classified into any of the three possibilities enumerated above, 
and we just ignore such processes by considering the extreme cases (i.e. modes deeply inside or outside the horizon).

\begin{table}[h]
\begin{center} 
\caption{Classification of the terms contributing to Eq.~\eqref{eq4-dNdt}.}
\vspace{3mm}
\begin{tabular}{c @{\hspace{5mm}} rcl}
\hline\hline
Case 1                            &                  & $k_H$ & $<k_1,k_2,k_3,p$\\
\hline
\multirow{5}{*}{Case 2 [$p>k_H$]} & $k_1, k_3 <$     & $k_H$ & $< k_2, p $\\
                                  & $k_1, k_2 <$     & $k_H$ & $< k_3, p$\\                                  & $k_3 <$          & $k_H$ & $< k_1, k_2, p$\\                                 & $k_1 <$          & $k_H$ & $ < k_2, k_3, p$\\                                  & $k_2 <$          & $k_H$ & $ < k_1, k_3, p$\\
\hline
\multirow{3}{*}{Case 2 [$p<k_H$]} & $p <$            & $k_H$ & $< k_1, k_2, k_3 $\\                                  & $k_3, p <$       & $k_H$ & $< k_1, k_2$\\
                                  & $k_2, p <$       & $k_H$ & $< k_1, k_3 $\\
                                  \hline 
Case 3                            & $k_1,k_2,k_3,p<$ & $k_H$ & \\
\hline
\multirow{6}{*}{Others}           
 			       & $k_2, k_3 <$     & $k_H$ & $< k_1, p$\\
                                   & $k_1, p <$       & $k_H$ & $< k_2, k_3$\\
                                  & $k_1, k_2, k_3<$ & $k_H$ & $< p$\\                                                                    & $k_1, k_2, p<$   & $k_H$ & $< k_3$\\                                         & $k_2, k_3, p<$   & $k_H$ & $< k_1$\\
                                  & $k_1, k_3, p<$   & $k_H$ & $< k_2$\\
                                 \hline\hline
\label{tab1}
\end{tabular}
\end{center}
\end{table}

Hereafter, we estimate the rate of processes contributing to the three cases considered above.
In order to estimate the interaction rate, we must evaluate the integration over three external momenta,
such as ${\bf k}_1$, ${\bf k}_2$, and ${\bf k}_3$ appearing on the right-hand side of Eq.~\eqref{eq4-dNdt_1}.
It is not straightforward to perform integration over these momentum variables
unless we fully specify the field dynamics beyond the horizon in the unitary gauge to know the exact spectrum of $\alpha_{\bf p}$.
Instead, we use a naive ``random walk" estimation~\cite{Erken:2011dz} by noting that the amplitude of $\alpha_{\bf p}$
is bounded by the number density of axions and that the value of $\alpha_{\bf p}$ varies randomly in the complex plane.
Strictly speaking, this assumption is only applicable to the case where PQ symmetry is broken after inflation,
while it is inapplicable to the case where PQ symmetry is broken before inflation.
In the following, we first consider the former case, and then discuss the latter case.

Let us discretize the integral in Eq.~\eqref{eq4-dNdt_1} as
\begin{equation}
\int\frac{d^3k}{(2\pi)^3}\bigg|_{k_H(t)<k\lesssim H_qa_q} \to \frac{1}{V}\sum_l^K, \nonumber\\
\end{equation}
where $l$ is an abbreviation for the label of the momentum ${\bf k}$, and $K$ represents the total number of states
between two scales $k_H(t)$ and $H_qa_q$.
Then, we can rewrite the right-hand side of Eq.~\eqref{eq4-dNdt_1} as
\begin{align}
&-4V\int\frac{d^3k_1}{(2\pi)^3}\bigg|_{k_H(t)<k_1\lesssim H_qa_q}\int\frac{d^3k_2}{(2\pi)^3}\bigg|_{k_H(t)<k_2\lesssim H_qa_q}F(t;|{\bf k}_1-{\bf p}|)
{\rm Im}(\alpha^*_{\bf k_1} \alpha^*_{\bf k_2} \alpha_{\bf k_1+k_2-p} \alpha_{\bf p}) \nonumber\\
&\quad \to \quad -\frac{4}{V}\sum_{l,m}^K{\rm Im}(\alpha^*_l\alpha^*_m\alpha_{l+m-p}\alpha_p)F(t;a\delta p), \label{eq4-19}
\end{align}
where $\delta p = |{\bf k}_1-{\bf p}|/a(t)$ is the typical momentum transferred in the process.
For the case where PQ symmetry is broken after inflation,
we assume that the coefficient $\alpha_l$ on the right-hand side of Eq.~\eqref{eq4-19}
varies randomly with the amplitude $|\alpha|\lesssim \sqrt{\mathcal{N}/K}$,
where $\mathcal{N}$ is the total number of axions and $\mathcal{N}/K$ represents
the average number of axions occupying the mode $l$ [see Eq.~\eqref{eq4-18}].
The summation in Eq.~\eqref{eq4-19} can be considered as a random walk in complex space with the number of steps $K$, and estimated as
\begin{equation}
\frac{1}{V}\sum_{l,m}^K{\rm Im}(\alpha^*_l\alpha^*_m\alpha_{l+m-p}\alpha_p)F(t;a\delta p) \sim \frac{\mathcal{N}^2}{VK}F(t;a\delta p) \sim \frac{\mathcal{N}}{K}na^3F(t;a\delta p), \label{eq4-20}
\end{equation}
where we use $\mathcal{N}/Va^3\sim n$, and $n$ is the number density of axions.
Applying this result to Eq.~\eqref{eq4-9} and ignoring a numerical factor of $\mathcal{O}(1)$, we obtain the estimation for the interaction rate
\begin{equation}
\Gamma \sim na^3 F(t;a\delta p). \label{eq4-21}
\end{equation}
Note that the value of $\Gamma$ simply depends on the structure of $F(t;a\delta p)$.
Though we have considered the right-hand side of Eq.~\eqref{eq4-dNdt_1} above (case 1),
a similar analysis can be applied to other cases to obtain Eq.~\eqref{eq4-21}.

Now that we have the formula for the interaction rate [Eq.~\eqref{eq4-21}], let us estimate it for individual cases.
First of all, consider case 1, where all modes contributing to the process
have wavelengths shorter than the horizon.
In this case we have $\delta p\gg k_H/a$,
except for the case where the momentum transfer vanishes,
and we can use the approximation
\begin{equation}
F(t;a\delta p) \to -\frac{2\pi Gm^2}{a^3(t)(\delta p)^2}, \label{eq4-limF}
\end{equation}
which holds for $\delta p\gg k_H/a$.
Here, we drop the highly oscillating terms [see Eq.~\eqref{fxlimit}].
Then, the interaction rate is estimated from Eq.~\eqref{eq4-21} as
\begin{equation}
\Gamma_1 \sim \frac{Gm^2 n}{(\delta p)^2}, \label{eq4-Gamma1}
\end{equation}
where the subscript ``1" represents the rate for the process classified to case 1.
It indeed reproduces the previous result [Eq.~\eqref{eq1-1}] derived on the basis of the Newtonian approximation.
Note that the dependence on the scale factor $\Gamma \propto a^{-1}$, which was introduced by hand
(i.e.~by extending the result obtained in the Minkowski background) in the previous works~\cite{Erken:2011dz,Saikawa:2012uk},
appeared automatically in the formalism used here.

Next, let us consider case 2. It should be noted that in this case we always have $\delta p\gg k_H/a$,
since either the external line labeled by $k_1$ ($k_2$) or that labeled by $p$ ($k_3$) is subhorizon sized.
Therefore, we can use the approximation in Eq.~\eqref{eq4-limF} and obtain
\begin{equation}
\Gamma_2 \sim \frac{Gm^2 n}{(\delta p)^2}, \label{eq4-Gamma2}
\end{equation}
where the subscript ``2" represents the rate for the processes classified to case 2.
This result implies that the transition between subhorizon modes and superhorizon modes occurs
rapidly in the same rate with $\Gamma_1$.

On the other hand, for case 3 we cannot use the approximation in Eq.~\eqref{eq4-limF},
since all external lines contributing to the process are superhorizon sized ($\delta p\ll k_H/a$).
We note that such a process is less relevant to the thermalization of the system, since
the thermalization is expected to proceed due to the transition into almost homogeneous (superhorizon-sized) modes
from modes with higher momentum (subhorizon-sized), or vice versa.
The contribution classified into case 3 seems to be interpreted differently from this transition process,
and we give some discussions on the rate for case 3 in Appendix~\ref{secC}.

Finally, we consider the case where PQ symmetry is broken before inflation.
In this case, we cannot apply the results obtained above, since the assumption on the amplitude of $\alpha_{\bf p}$ becomes different.
As was mentioned before, the typical amplitude of $\alpha_{\bf p\ne 0}$ is suppressed by a factor of $\mathcal{O}(10^{-5})$
compared with that of $\alpha_{\bf p=0}$.
In other words, the fraction of the number of axions occupying the state with finite momentum  $p$ in the phase space is less than $\mathcal{O}(10^{-10})$.
For case 1, we can still use the random walk estimation by assuming that
$\alpha_{\bf p\ne 0}$ varies randomly with the amplitude $|\alpha|\lesssim\mathcal{O}(10^{-5})\times \sqrt{\mathcal{N}/K}$ in Eq.~\eqref{eq4-19}, and obtain
\begin{equation}
\Gamma_1 \sim \mathcal{O}(10^{-10})\times\frac{Gm^2 n}{(\delta p)^2}.
\end{equation}
Since there is a severe suppression factor, we expect that the process classified to case 1 remains irrelevant, at least during the radiation-dominated era.
On the other hand, for case 2 (with $p>k_H$), the leading contribution in Eq.~\eqref{eq4-19} comes from the term in which two $\alpha$'s correspond to
the background value $|\alpha|\sim\sqrt{\mathcal{N}}$ and the other two $\alpha$'s correspond to the small fluctuations $|\alpha|\lesssim\mathcal{O}(10^{-5})\times \sqrt{\mathcal{N}/K}$.
Then Eq.~\eqref{eq4-20} is replaced by
\begin{equation}
\frac{1}{V}\sum_{l,m}{\rm Im}(\alpha^*_l\alpha^*_m\alpha_{l+m-p}\alpha_p)F(t;a\delta p) \sim \mathcal{O}(10^{-10})\times\frac{\mathcal{N}^2}{VK}F(t;a\delta p) \sim \mathcal{O}(10^{-10})\times\frac{\mathcal{N}}{K}na^3F(t;a\delta p)
\end{equation}
Dividing it by $\langle\mathcal{N}_{\bf p}\rangle = |\alpha_{\bf p}|^2\sim \mathcal{O}(10^{-10})\times(\mathcal{N}/K)$, we obtain the same results as in Eq.~\eqref{eq4-Gamma2}:
\begin{equation}
\Gamma_2 \sim \frac{Gm^2 n}{(\delta p)^2}.
\end{equation}
This result implies that, in the case where PQ symmetry is broken before inflation, only the transition between a large homogeneous mode and modes with higher momentum 
occurs rapidly, while transitions among modes with higher momentum are highly suppressed.

In summary, we see that the interaction rate for the processes involving the modes inside the horizon
agrees with the expression in Eq.~\eqref{eq1-1} obtained from the Newtonian approximation,
but the classification of the processes becomes different according to whether PQ symmetry is broken before inflation or not.
If PQ symmetry is broken after inflation, both $\Gamma_1$ and $\Gamma_2$ contribute to the right-hand side of Eq.~\eqref{eq4-dNdt},
while $\Gamma_1$ is suppressed by a factor of $\mathcal{O}(10^{-10})$ compared with $\Gamma_2$ if PQ symmetry is broken before inflation.
What is notable is that in both cases,
the rate of the process corresponding to the transition between modes inside and outside the horizon (case 2)
is estimated by Eq.~\eqref{eq1-1}.
We interpret this contribution as the process where the modes with short wavelength fall into those with long wavelength by exchanging gravitons with short wavelength.
Although the homogeneous mode does not feel the gravitational force, it is possible for the modes with short wavelength to annihilate and produce a homogeneous wave.
In this sense, this process can be regarded as a causal one, which is mediated by the modes whose wavelength is shorter than the horizon,
and hence there is a possibility for it to have relevance to the gravitational thermalization.

\section{\label{sec5} Summary and discussions}
In this work, we considered gravitational interactions of coherently oscillating axions in detail.
Action for the system of a general massive scalar field in the FRW background was constructed, and the interaction Hamiltonian
for the scalar-graviton system was explicitly obtained.
The quartic interaction between a massive scalar field can be understood as the process mediated by the dynamical field $\zeta$,
which is interpreted as a fluctuation of the background fluids such as radiations.
By integrating out the gravitational degrees of freedom, the effective Hamiltonian for the quartic interaction [Eq.~\eqref{total_Heff}], which can be used in the 
tree level calculations, is also derived.
Using the effective interaction Hamiltonian, we estimated the interaction rate of dark matter axions in coherent states.
The interaction processes can be classified into three cases: the process involving modes inside the horizon (case 1),
that accompanied by a transition between modes inside and outside the horizon (case 2),
and that involving modes outside the horizon (case 3).
We found that the interaction rate for cases 1 and 2 reproduces previous estimation obtained from the Newtonian approximation
if PQ symmetry is broken after inflation, while only case 2 is relevant if PQ symmetry is broken before inflation.
In particular, it turned out that the processes classified into case 2 can occur rapidly, and these are presumed to be relevant to the formation of BEC.

The interaction rate given by Eq.~\eqref{eq4-Gamma1} or Eq.~\eqref{eq4-Gamma2} obeys the scaling
$\Gamma\propto a^{-1}$, and it eventually exceeds the expansion rate when the temperature of the universe becomes
$T\sim \mathrm{keV}$~\cite{Erken:2011dz,Saikawa:2012uk}.
However, it is still nontrivial whether the thermalization occurs at that time.
We note that our formalism is based on the quantum field theory, in which the time evolution is unitary, and hence the increase of the entropy cannot be discussed unless we introduce
some measure of coarse graining.
We just showed that the distribution function can change in the time scale given by Eq.~\eqref{eq4-Gamma1} or Eq.~\eqref{eq4-Gamma2}, but we did not show how it evolves with time.
If the gravitational thermalization actually occurs, we expect that the axion field is somewhat homogenized in the position space.
The problem is whether such a dissipative effect exists in the system.
A careful estimation based on the linearized Einstein equations indicates that the damping scale of the axion density fluctuation is not long enough to homogenize the axion field~\cite{Davidson:2013aba}.
Perhaps it is necessary to identify the components that are relevant to the gravitational dissipation in the system correctly, and to solve the axion evolution explicitly in order to clarify these issues.

Finally, let us comment on the limitations on the use of the formalism developed in this work.
In this paper, gravitational interactions of axions are formulated by the use of the notion of EFT.
In the unitary gauge, it is possible to discuss gravitational self-interactions of axions unambiguously, since we can treat the evolution of the background field, axions, and gravitons separately.
This formalism is applicable to the system during the radiation-dominated era regardless of whether PQ symmetry is broken before inflation or not, as long as the effects of the axion field on the background dynamics are negligible.
However, it should be noted that this formalism will not be applicable to the system in the matter-dominated era if PQ symmetry is broken after inflation.
In the matter-dominated (axion-dominated) era, the time evolution of the axion field itself determines the evolution of the background field.
Then, if PQ symmetry is broken after inflation, we cannot use the decomposition into the smooth background field and small fluctuations.
In other words, the clock of the universe differs for each horizon, and spatial diffeomorphisms are also broken.
Since there are few symmetries to specify the form of interactions, it is a tough problem to write down the possible gravitational interactions in such a system.
Fortunately, the gravitational thermalization (if it truly occurs) begins in the radiation-dominated era at $T\sim \mathrm{keV}$, where our formalism is still applicable.
We also note that we can use this EFT approach in the matter-dominated era for the case where PQ symmetry is broken before inflation, since in that case
it is possible to define the background axion field which only breaks time diffeomorphisms.
\begin{acknowledgments}
The authors have greatly benefited from discussions with Davidson Sacha, Martin Elmer and Georg G. Raffelt.
T.~N.~is supported
by Special Postdoctoral Researchers
Program at RIKEN.
K.~S.~and R.~S.~are supported by the Japan Society for the Promotion of Science (JSPS) through research fellowships.
The work of M.~Y.~is supported in part by the Grant-in-Aid for Scientific Research on Innovative Areas No.~24111706
and the Grant-in-Aid for Scientific Research No.~25287054.
\end{acknowledgments}

\appendix
\section{Details of the tensor calculations}
\label{secA}
In this appendix, we calculate quantities
which are used in the derivation of the action in Eqs.~\eqref{deltag_second},~\eqref{phi_cubic}.
First,
the three-dimensional Ricci scalar is given by
\begin{align}
R^{(3)} &= h^{ij}R^{(3)}_{ij} \nonumber\\
&=h^{ij}\left(\partial_k\Gamma^{(3)k}_{\quad ij}-\partial_j\Gamma^{(3)k}_{\quad ki}+\Gamma^{(3)k}_{\quad km}\Gamma^{(3)m}_{\quad ij}-\Gamma^{(3)k}_{\quad mi}\Gamma^{(3)m}_{\quad kj}\right) \nonumber\\
\nonumber
&=\partial_i\partial_j h_{kl}(h^{ik}h^{jl} - h^{ij}h^{kl})\\
&\quad
+\frac{1}{4}\partial_i h_{jk}\partial_l h_{mn}(
3h^{il}h^{jm}h^{kn}
-h^{il}h^{jk}h^{mn}
+4h^{ij}h^{kl}h^{mn}
-4h^{ij}h^{km}h^{ln}
-2h^{im}h^{jl}h^{kn})\,,\label{eqA-1}
\end{align}
where $R^{(3)}_{ij}$ and $\Gamma^{(3)k}_{\quad ij}$ are the three-dimensional Ricci tensor and Christoffel symbol, respectively.
In the unitary gauge 
with the transverse conditions,
the spatial metric and its inverse are given by
\begin{equation}
h_{ij} = a^2e^{2\zeta}(e^{\gamma})_{ij}, \quad h^{ij} = a^{-2}e^{-2\zeta}(e^{-\gamma})_{ij}
\quad
{\rm with}
\quad
\partial_i\gamma_{ij}=0. \label{eqA-2}
\end{equation}
Substituting them into Eq.~\eqref{eqA-1}, we have
\begin{align}
\nonumber
R^{(3)}&=a^{-2}e^{-2\zeta}\Big[
-4\partial_i\partial_j\zeta(e^{-\gamma})_{ij}
-2\partial_i\zeta\partial_j\zeta(e^{-\gamma})_{ij}
-4\partial_i\zeta(\partial_j e^{-\gamma})_{ij}
+\frac{1}{4}(e^{-\gamma})_{ij}(\partial_i e^{-\gamma}\partial_j e^{\gamma})_{kk}\\
\nonumber
&\quad\qquad\qquad
+(e^{-\gamma}\partial_i\partial_j e^\gamma\,e^{-\gamma})_{ij}
+(\partial_i e^{-\gamma}\partial_j e^\gamma\, e^{-\gamma})_{ij}
+\frac{1}{2}(\partial_i e^{-\gamma}\partial_j e^\gamma\, e^{-\gamma})_{ji}
\Big]\\
&=a^{-2}e^{-2\zeta}\Big[
-4\partial^2\zeta-2(\partial_i\zeta)^2
-\frac{1}{4}\partial_k\gamma_{ij}\partial_k\gamma_{ij}+\ldots\Big]\,, \label{eqA-3}\\
\sqrt{h}R^{(3)}&=a\Big[
-4\partial^2\zeta-4\zeta\partial^2\zeta-2(\partial_i\zeta)^2
-\frac{1}{4}\partial_k\gamma_{ij}\partial_k\gamma_{ij}+\ldots\Big]\,, \label{eqA-4}
\end{align}
where dots represents terms of at least third order in $\zeta$ and $\gamma$.

Next, let us compute the quantity
\begin{align}
E_{ij}&=\frac{1}{2}\left(\dot{h}_{ij}-\nabla^{(3)}_i N_j-\nabla^{(3)}_j N_i\right)\,. \label{eqA-5}
\end{align}
Note that
\begin{align}
\nonumber
E_{i}^j&=h^{jk}E_{ik}\\
\nonumber
&=(H+\dot{\zeta})\delta_i^j+\frac{1}{2}(e^{-\gamma}\partial_te^\gamma)_{ji}
-\frac{1}{2}\left(\partial_i N^j+h^{jl}h_{ik}\partial_l N^k
+h^{jl}\partial_m h_{li}N^m
\right)\\
&=\frac{1}{2}h^{jk}\partial_th_{ki}
-\frac{1}{2}\left(\partial_i N^j+h^{jl}h_{ik}\partial_l N^k
+h^{jl}\partial_m h_{li}N^m
\right)\,, \label{eqA-6}
\\
\nonumber
E&=E_i^i\\
&=3(H+\dot{\zeta})
-\partial_i N^i
-3\partial_m\zeta N^m\,.\label{eqA-7}
\end{align}
We therefore have
\begin{align}
\nonumber
E_{ij}E^{ij}-E^2&=-6H^2-12H\dot{\zeta}
+4H\partial_iN^i
-6\dot{\zeta}^2
+4\dot{\zeta}\partial_iN^i
-(\partial_iN^i)^2
+12H\partial_m\zeta N^m\\
&\quad
+\frac{1}{4}\dot{\gamma}_{ij}\dot{\gamma}_{ij}
-\dot{\gamma}_{ij}\partial_iN^j
+\frac{1}{2}(\partial_j N^i)(\partial_i N^j)
+\frac{1}{2}(\partial_j N^i)(\partial_j N^i)
+\ldots\,. \label{eqA-8}
\end{align}
Expansions given in the above equations are used
to obtain the expressions in Eqs.~\eqref{deltag_second} and \eqref{phi_cubic}.

\section{Mode expansion in the interaction picture}
\label{secB}
In this appendix,
we summarize
mode expansions
of the fields $\phi$ and $\zeta$
in the interaction picture.
Here we consider the system whose Hamiltonian is given by
\begin{align}
H[\phi_a,\pi_a;t] &=  H_{\rm free}[\phi_a,\pi_a;t] + H_{\rm int}[\phi_a,\pi_a;t], \label{eqB-1-1}\\
H_{\rm free}[\phi_a,\pi_a;t] &= \int d^3x\mathcal{H}_{\rm free}, \label{eqB-1-2} \\
H_{\rm int}[\phi_a,\pi_a;t] &= \int d^3x\mathcal{H}_{\rm int}, \label{eqB-1-3}
\end{align}
where $\mathcal{H}_{\rm free}$ and $\mathcal{H}_{\rm int}$ are given by Eq.~\eqref{eq3-12} and Eq.~\eqref{eq3-13}, respectively.
Note that the Hamiltonian $H[\phi_a,\pi_a;t]$ explicitly depends on time $t$, which is due to the time evolution of the background.
In such a system, the time evolution of the expectation value of an operator $\mathcal{O}$ constructed from $\phi_a$ and $\pi_a$ is given by~\cite{Weinberg:2005vy}
\begin{align}
\nonumber
\langle\mathrm{in}|\mathcal{O}(t)|\mathrm{in}\rangle
&=\langle{\rm in}|
\left[\bar{T}\exp \Big(i\int_{t_0}^tdt^\prime H_{\rm int}(t^\prime)\Big)\right]
\mathcal{O}^I(t)
\left[T\exp \Big(-i\int_{t_0}^tdt^\prime H_{\rm int}(t^\prime)\Big)\right]
|{\rm in}\rangle\\
&= \sum^{\infty}_{N=0}i^N\int^t_{t_0}dt_N\int^{t_N}_{t_0}dt_{N-1}\dots\int^{t_2}_{t_0}dt_1
\langle{\rm in}|[H_{\rm int}(t_1),[H_{\rm int}(t_2),\dots[H_{\rm int}(t_N),{\cal O}^I(t)]\dots]]
|{\rm in}\rangle, \label{eqB-1-4}
\end{align}
where $|\mathrm{in}\rangle$ is an ``in-state" specified at some initial time $t_0$.
On the right-hand side of Eq.~\eqref{eqB-1-4}, $H_{\rm int}(t)$ and ${\cal O}^I(t)$ are constructed from interaction picture fields,
which satisfy the free-field equations
\begin{align}
\dot{\phi}_a({\bf x}, t) = i \left[H_{\rm free}[\phi_a,\pi_a], \phi_a ({\bf x}, t)\right], \label{eqB-1-5}\\
\dot{\pi}_a({\bf x}, t) = i \left[H_{\rm free}[\phi_a,\pi_a], \pi_a ({\bf x}, t)\right], \label{eqB-1-6}
\end{align}
with
\begin{align}
&\left[\phi_a ({\bf x}, t), \pi_b ({\bf y}, t)\right] = i\delta_{ab}\delta^{(3)}({\bf x-y}), \nonumber\\
&\left[\phi_a ({\bf x}, t), \phi_b ({\bf y}, t)\right] = \left[\pi_a ({\bf x}, t), \pi_b ({\bf y}, t)\right] = 0. \label{eqB-1-7}
\end{align}

Now, let us take the mode expansion of the interaction picture fields.
Since the relevant terms for the interaction Hamiltonian are couplings of the form $\phi^4$ and $\zeta\phi^2$ (see discussions in Sec.~\ref{sec3-1}),
for our purpose it is enough to consider the fields $\phi$ and $\zeta$ only.
For the scalar field $\phi$, Eqs.~\eqref{eqB-1-5} and \eqref{eqB-1-6} lead to
\begin{equation}
\dot{\phi} = a^{-3}\pi_{\phi}, \qquad
\dot{\pi}_{\phi} = a^3\left[a^{-2}\partial^2\phi - m^2 \phi\right]. \label{eqB-1-8}
\end{equation}
Hence, $\phi$ can be expanded as
\begin{equation}
\phi({\bf x},t) = \int \frac{d^3 k}{(2\pi)^3}\left[e^{i{\bf k\cdot x}}\varphi_k(t)a_{\bf k} + e^{-i{\bf k\cdot x}}\varphi^*_k(t)a^{\dagger}_{\bf k} \right], \label{eqB-1-9}
\end{equation}
where $a_{\bf k}$ is the annihilation operator satisfying the usual commutation relations
\begin{align}
&[a_{\bf k}, a^{\dagger}_{\bf k'}] = (2\pi)^3\delta^{(3)}(\bf{k-k'}), \nonumber\\
&[a_{\bf k}, a_{\bf k'}] = [a^{\dagger}_{\bf k}, a^{\dagger}_{\bf k'}] = 0, \label{eqB-1-10}
\end{align}
and $\varphi_k(t)$ is the positive-frequency solution of the free-field equation in $k$ space
\begin{equation}
\ddot{\varphi}_k + 3H\dot{\varphi}_k + \left(m^2+\frac{k^2}{a^2}\right)\varphi_k = 0, \label{eqB-1-11}
\end{equation}
where $k\equiv|{\bf k}|$.
Assuming the power-law expansion 
$a\propto t^{\beta}$,
we reduce Eq.~\eqref{eqB-1-11} into the form
\begin{equation}
\ddot{u}_k(t) + \omega_k^2(t)u_k(t) = 0, \label{eqB-1-12}
\end{equation}
where $\varphi_k = a^{-3/2}u_k$, and
\begin{equation}
\omega_k^2(t) = \frac{9}{4}\beta\left(\frac{2}{3}-\beta\right)t^{-2} + m^2 + \frac{k^2}{a^2}. \label{eqB-1-13}
\end{equation}
As long as $\dot{\omega}_k/\omega_k \ll \omega_k$ is satisfied,
we can use the WKB solution\footnote{
It would be notable that
the equation of motion [Eq.~\eqref{eqB-1-11}]
can be solved analytically for the radiation-dominated era,
although it is difficult to solve Eq.~\eqref{eqB-1-11} in general.
Using $a\propto t^{1/2}$,
we obtain the following form of the mode function
$\varphi_k(t)$:
\begin{align}
\label{eqB-1-14}
\varphi_k(t)&=\frac{t^{3/4}}{a^{3/2}}\frac{2^{\frac{3}{4}+i\frac{(k/a)^2}{2m}t}(im)^{\frac{3}{4}+i\frac{(k/a)^2}{2m}t}}{\sqrt{2m}}
e^{-imt}\,U\Big(\frac{3}{4}+i\frac{(k/a)^2}{2m}t,\frac{3}{2},2imt\Big)\,,
\end{align}
where
$U(a,b,z)$ is Tricomi's confluent hypergeometric function
and the mode function $\varphi_k(t)$ is normalized such that $a^3(\varphi_k\dot{\varphi}_k^\ast-\dot{\varphi}_k\varphi_k^\ast)=i$.
In the late-time limit $t\to\infty$,
it reproduces the WKB solution [Eq.~\eqref{eqB-1-16}]:
\begin{align}
\varphi_k(t)&\simeq\frac{a^{-3/2}}{\sqrt{2m}}
\exp \left[-imt\Big(1+(\ln a)\frac{(k/a)^2}{m^2}\Big)\right]
\simeq\frac{a^{-3/2}}{\sqrt{2m}}e^{-imt}\,, \label{eqB-1-15}
\end{align}
where we have used
$\displaystyle\lim_{t\to\infty}a(t)=\infty$
at the second equality.
Although we have an analytic expression in Eq.~\eqref{eqB-1-14}
for the mode function $\varphi_k$,
we use the WKB solution [Eq.~\eqref{eqB-1-16}]
in Secs.~\ref{sec3} and~\ref{sec4}
because its use is justified
in the regime $m\gg H,k/a$.}
\begin{equation}
\varphi_k(t) \simeq \frac{a^{-3/2}}{\sqrt{2\omega_k(t)}}\exp\left(-i\int^t\omega_k(t')dt'\right). \label{eqB-1-16}
\end{equation}
Note that $\omega \simeq m$ for $m\gg H,k/a$,
which is used in Secs.~\ref{sec3} and~\ref{sec4}.

Similarly, for $\zeta$ we have
\begin{equation}
\zeta({\bf x},t) = \int \frac{d^3 k}{(2\pi)^3}\left[e^{i{\bf k\cdot x}}{\cal Z}_k(t)a_{z{\bf k}} + e^{-i{\bf k\cdot x}}{\cal Z}^*_k(t)a^{\dagger}_{z{\bf k}} \right], \label{eqB-1-17}
\end{equation}
where $a_{z{\bf k}}$ is the annihilation operator satisfying the commutation relations
\begin{align}
&[a_{z{\bf k}}, a^{\dagger}_{z{\bf k'}}] = (2\pi)^3\delta^{(3)}(\bf{k-k'}), \nonumber\\
&[a_{z{\bf k}}, a_{z{\bf k'}}] = [a^{\dagger}_{z{\bf k}}, a^{\dagger}_{z{\bf k'}}] = 0, \label{eqB-1-18}
\end{align}
and ${\cal Z}_k(t)$ is the positive-frequency solution of the following equation:
\begin{equation}
\ddot{\cal Z}_k +\left(3H+\frac{\ddot{H}}{\dot{H}}-2\frac{\dot{H}}{H}-2\frac{\dot{c}_s}{c_s}\right) \dot{\cal Z}_k +c_s^2\frac{k^2}{a^2}{\cal Z}_k = 0. \label{eqB-1-19}
\end{equation}
When the background expands
with a power-law $a\propto t^{\beta}$
and the time dependence of the sound speed $c_s$ is negligible,
this equation is exactly solved, and we obtain
\begin{equation}
{\cal Z}_k(t) = M_{\rm Pl}^{-1}\sqrt{\frac{\beta\pi\tau}{2}}\frac{c_s}{2a}\exp\left[-i\frac{\pi}{2}\left(\nu+\frac{1}{2}\right)\right]H^{(2)}_{\nu}(c_s k\tau),
\label{eqB-1-20}
\end{equation}
where $H_{\nu}^{(2)}(x) = J_{\nu}(x)-iN_{\nu}(x)$ is the Hankel function of the second kind with $\nu = (3\beta -1)/2(1-\beta)$,
and $\tau$ is conformal time given by $d\tau=dt/a$.
The normalization of the mode function in Eq.~\eqref{eqB-1-20} is determined from the requirement that
the solution match the
positive frequency solution
in the limit $\tau\to \infty$\footnote{This choice of the mode function [and hence the choice of the vacuum $|0\rangle$ in Eq.~\eqref{eq4-def_of_vac}] corresponds to 
what is called the adiabatic vacuum~\cite{Birrell:1982ix},
on which a comoving observer fails to detect particle number for large $k$ (in particular $k\tau \gg 1$) at any $\tau$.
This implies that the state $|0\rangle$ can be interpreted as a vacuum in a good approximation,
at least for the modes inside the horizon every time in the
power-law expanding phase.}:
\begin{equation}
{\cal Z}_k(t) \xrightarrow{\tau \to \infty} (\sqrt{2} M_{\rm Pl})^{-1}\frac{\sqrt{\beta c_s}}{\sqrt{2k}a}\exp(-ic_s k\tau). \label{eqB-1-21}
\end{equation}
In the radiation-dominated universe with $\beta=\nu=1/2$, the mode function in Eq.~\eqref{eqB-1-20} reduces to
\begin{equation}
{\cal Z}_k(t) = \frac{\sqrt{\pi\tau}c_s}{4aM_{\rm Pl}}(-i)H_{1/2}^{(2)}(c_s k\tau) = \frac{c_s}{2\sqrt{2}aM_{\rm Pl}}\frac{1}{\sqrt{c_s k}}e^{-ic_s k\tau}. \label{eqB-1-22}
\end{equation}


\section{Interaction rate for the modes outside the horizon}
\label{secC}
In Sec.~\ref{sec4}, we classified the quartic interaction processes into three cases according to whether the wavelength of the modes
contributing to them is shorter than the (sound) horizon or not.
Among them, we estimated the interaction rates $\Gamma_1$ and $\Gamma_2$ for the processes involving modes inside the horizon (cases 1 and 2),
which are expected to be relevant to the gravitational thermalization.
On the other hand, the process classified into case 3 cannot be interpreted as a causal interaction mediated by waves inside the horizon.
In this appendix, we attempt to give some interpretation of the result obtained for this exceptional process.

First of all, let us estimate the interaction rate for case 3 by using Eq.~\eqref{eq4-21}.
Since all external lines attached to the quartic coupling have wavelength longer than the horizon ($\delta p\ll k_H/a$),
the factor $F(t;a\delta p)$ can be approximated as
\begin{equation}
F(t;a\delta p) \to \frac{\pi Gm^2c_s^2}{a^3(t)H^2(t)} + \mathcal{O}\left(\left(\frac{\delta p}{k_H/a}\right)^2\right). \label{eqC-1}
\end{equation}
According to Eq.~\eqref{eq4-21}, this leads to the estimation for the interaction rate
\begin{equation}
\Gamma_3 \sim \frac{Gm^2n}{H^2}, \label{eqC-2}
\end{equation}
where the subscript ``3" represents the rate for the process classified to case 3.
In the radiation-dominated era, this quantity evolves as $\Gamma_3 \propto a$,
which is faster than the scaling $\Gamma_{1,2} \propto a^{-1}$ for the interaction rates given by Eqs.~\eqref{eq4-Gamma1} and~\eqref{eq4-Gamma2}.

Although the result in Eq.~\eqref{eqC-2} implies that the rate of the process  
for the modes outside the horizon is faster than that of the processes accompanied by the modes inside the horizon $\Gamma_{1,2}$,
we interpret that this contribution is not relevant to the thermalization,
since they represent the evolution of the modes outside the horizon and are not regarded as the causal process.
Rather, it would be appropriate to interpret that the result [Eq.~\eqref{eqC-2}] just represents the evolution rate of the homogenous background field.
Here, we note that the distinction between nonzero modes $|\alpha_{\bf p}\rangle$ with $p<k_H$
and the exact zero mode $|\alpha_{\bf p=0}\rangle$ becomes ambiguous, since the uncertainty principle implies that the wave with $p<k_H$
and the homogeneous field
are indistinguishable within the cosmological time scale.
The homogeneous background field does evolve as long as some interaction terms exist in the action.
In the system considered here, the evolution of the homogeneous background field is affected by the effective quartic interaction given by Eq.~\eqref{eq3-3-Heffout}.

The emergence of the quartic interaction acting on the homogeneous field can be understood when we consider the quantum corrections to
the evolution of the classical background field in $\phi$.
Let us start with the action given by Eqs.~\eqref{eq2-3-12}-\eqref{eq2-3-19}, and consider the construction of the effective action $\Gamma_{\rm eff}[\phi_{\rm cl}]$
for the classical background field $\phi_{\rm cl}$, which is equivalent to $\langle\alpha_{{\bf p}={\bf 0}}|\phi|\alpha_{{\bf p}={\bf 0}}\rangle$.
As described in standard textbooks, the effective action $\Gamma_{\rm eff}[\phi_{\rm cl}]$ can be obtained by integrating out all of the quantum fluctuations around
the background field $\phi_{\rm cl}$ in the path integrals
\begin{equation}
e^{i\Gamma_{\rm eff}[\phi_{\rm cl}]} = \int {\cal D}\phi'{\cal D}\zeta{\cal D}\gamma_{ij} e^{iS[\phi',\zeta,\gamma_{ij};\phi_{\rm cl}]}, \label{eqC-3}
\end{equation}
where $S[\phi',\zeta,\gamma_{ij};\phi_{\rm cl}]$ represents the action obtained by substituting $\phi\to\phi_{\rm cl}+\phi'$
into Eq.~\eqref{eq2-3-12}, and $\phi'$ is the quantum fluctuation around the classical value $\phi_{\rm cl}$.
Applying the similar discussion with Sec.~\ref{sec3-1}, we deduce that the leading contribution to the scalar quartic interaction
is given by the second term of Eq.~\eqref{eq2-3-18},
which is proportional to $\dot{\zeta}(\dot{\phi}^2-m^2\phi^2)$.
Hence, it will be enough to consider the following action to see the minimal effect arising from the quantum corrections:
\begin{align}
S = \int d^4 x a^3\Bigg[ M_{\rm Pl}^2\tilde{\epsilon}\left(\dot{\zeta}^2-c_s^2\frac{(\partial_i\zeta)^2}{a^2}\right)
+\frac{1}{2}\left(\dot{\phi}^2  -m^2\phi^2\right)
-\frac{1}{2H}\dot{\zeta}\left(\dot{\phi}^2 + m^2\phi^2\right)\Bigg]. \label{eqC-4}
\end{align}
Note that this action can be deformed as follows:
\begin{align}
S &= \int d^4 x a^3\Bigg[ M_{\rm Pl}^2\tilde{\epsilon}\left(\dot{\zeta}-\frac{1}{4HM_{\rm Pl}^2\tilde{\epsilon}}
\left(\dot{\phi}^2+m^2\phi^2\right)\right)^2
+\frac{1}{2}\left(\dot{\phi}^2  -m^2\phi^2\right)
-\frac{1}{16H^2M_{\rm Pl}^2\tilde{\epsilon}}\left(\dot{\phi}^2  + m^2\phi^2\right)^2\Bigg], \label{eqC-5}
\end{align}
where we ignore the term containing the spatial derivative of $\zeta$, since we are considering the modes outside the horizon.
Then, integrating out the fluctuation $\zeta$, we are left with the following effective action:
\begin{equation}
\Gamma_{\rm eff}[\phi_{\rm cl}] = \int d^4x a^3\left[\frac{1}{2}\dot{\phi}_{\rm cl}^2 - \frac{1}{2}m^2\phi_{\rm cl}^2 
-\frac{1}{16H^2M_{\rm Pl}^2\tilde{\epsilon}}\left(\dot{\phi}_{\rm cl}^2 + m^2\phi_{\rm cl}^2\right)^2+\dots\right], \label{eqC-6}
\end{equation}
where the dots correspond to the loop corrections.
After varying $\Gamma_{\rm eff}[\phi_{\rm cl}]$, we obtain the evolution equation for the background field,
\begin{equation}
\ddot{\phi}_{\rm cl} + 3H\dot{\phi}_{\rm cl}+m^2\phi_{\rm cl} = -\frac{c_s^2m^2}{8H^2M_{\rm Pl}^2}\phi_{\rm cl}\left(\dot{\phi}_{\rm cl}^2 + m^2\phi_{\rm cl}^2\right) 
+ \frac{c_s^2}{8M_{\rm Pl}^2a^3}\frac{d}{dt}\left[\frac{a^3}{H^2}\dot{\phi}_{\rm cl}\left(\dot{\phi}_{\rm cl}^2 + m^2\phi_{\rm cl}^2\right)\right], \label{eqC-7}
\end{equation}
where we used $\tilde{\epsilon}=2c_s^{-2}$, which holds in the radiation-dominated universe.
Since we are considering the regime $m\gg H$, we can use the approximation $\ddot{\phi}\approx -m^2\phi$ on the right-hand side to obtain
\begin{equation}
\ddot{\phi}_{\rm cl} + 3H\dot{\phi}_{\rm cl}+m^2\phi_{\rm cl} \approx -\frac{c_s^2m^2}{4H^2M_{\rm Pl}^2}\phi_{\rm cl}\left(\dot{\phi}_{\rm cl}^2 + m^2\phi_{\rm cl}^2\right). \label{eqC-8}
\end{equation}

The result shown in Eq.~\eqref{eqC-2} is recast by means of
the evolution of the classical background field obeying Eq.~\eqref{eqC-8}.
Since the number operator $\mathcal{N}_{\bf p}(t)$ is constructed from the creation and annihilation operators
which diagonalize the free part of the Hamiltonian, 
in our calculation based on the in-in formalism
we expect that the quantity $\langle\mathcal{N}_{\bf p}(t)\rangle$
in part includes the evolution of the comoving ``number" $N_{\rm free}$, which consists of the free part
of the energy density of the background field:
\begin{equation}
N_{\rm free} = \frac{a^3\rho_{\rm free}}{m} \qquad {\rm with} \qquad \rho_{\rm free} = \frac{1}{2}\dot{\phi}_{\rm cl}^2 + \frac{1}{2}m^2\phi_{\rm cl}^2. \label{eqC-9}
\end{equation}
For such a quantity, we obtain the relation
\begin{equation}
\dot{\rho}_{\rm free} \approx -3H\dot{\phi}_{\rm cl}^2 -\frac{c_s^2m^2}{4H^2M_{\rm Pl}^2}\phi_{\rm cl}\dot{\phi}_{\rm cl}\left(\dot{\phi}_{\rm cl}^2 + m^2\phi_{\rm cl}^2\right) \label{eqC-10}
\end{equation}
from the equation of motion for the background field [Eq.~\eqref{eqC-8}].
The relation in Eq.~\eqref{eqC-10} leads to the fact that
\begin{equation}
\frac{\dot{N}_{\rm free}}{N_{\rm free}} \sim - \frac{c_s^2m^3\phi_{\rm cl}^2}{2H^2M_{\rm Pl}^2}, \label{eqC-11}
\end{equation}
where we use the naive estimation $\dot{\phi}_{\rm cl}\sim m\phi_{\rm cl}$.
This quantity coincides with what we estimated in Eq.~\eqref{eqC-2}, since the number density of axions is given by $n\sim m\phi_{\rm cl}^2$.

Equations~\eqref{eqC-10} and~\eqref{eqC-11} just imply that the value of the homogeneous field $\phi_{\rm cl}(t)$ changes
due to the dynamics driven by the effective action [Eq.~\eqref{eqC-6}].
We do not interpret that axions change the number in the time scale shown in Eq.~\eqref{eqC-2}, since the second term of the right-hand side of
Eq.~\eqref{eqC-10} merely corresponds to the small correction to the
evolution of the background field,
rather than the effect of the causal process induced by the interactions of waves inside the horizon.
It should be emphasized that the existence of this correction term itself
does not lead to any modification to the usual result for the evolution of the axion field,
since it remains smaller than the mass term $m^2\phi_{\rm cl}^2/2$ during the radiation-dominated era, where our formalism based on EFT remains applicable.


\end{document}